\documentclass[article,10pt]{IEEEtran}
\usepackage[utf8]{inputenc}
\usepackage{graphics}
\usepackage{graphicx}
\usepackage{stfloats}
\usepackage{amssymb}
\usepackage{amsmath}
\usepackage{amsfonts}
\usepackage{subfigure}
\usepackage{float}
\usepackage{pifont}
\usepackage{setspace}
\usepackage{multirow}
\usepackage{hyperref}
\usepackage{color}
\usepackage{url}
\usepackage[named]{algo}
\usepackage{hyperref}

\definecolor{light}{gray}{.93}

\def\QEDclosed{\mbox{\rule[0pt]{1.3ex}{1.3ex}}} 

\def\QED{\QEDclosed} 
\def\proof{\noindent\hspace{0em}{\itshape Proof: }}
\def\endproof{\hspace*{\fill}~\QED\par\endtrivlist\unskip}

\newtheorem{theorem}{Theorem}
\newtheorem{lemma}{Lemma}

\def\btheta{\boldsymbol{\theta}}
\def\bone{\mbox{\boldmath$1$}}
\def\bzero{\mbox{\boldmath$0$}}

\def\urltilda{\kern -.15em\lower .7ex\hbox{\~{}}\kern .04em}
\def\urldot{\kern -.10em.\kern -.10em}
\def\urlhttp{http\kern -.10em\lower -.1ex\hbox{:}\kern -.12em\lower 0ex\hbox{/}\kern -.18em\lower 0ex\hbox{/}}

\DeclareMathOperator{\sign}{sign}
\DeclareMathOperator{\prox}{prox}
\DeclareMathOperator{\proj}{proj}
\DeclareMathOperator{\conv}{conv}

\begin{document}

\title{The Ordered Weighted $\ell_1$ Norm: Atomic Formulation, Projections, and Algorithms}
\author{Xiangrong Zeng, \emph{Student Member, IEEE},  and   M\'{a}rio A. T. Figueiredo, \emph{Fellow, IEEE}  \thanks{Manuscript submitted on December ??, 20??. } \\
\thanks{ Both authors are with the Instituto de Telecomunica\c{c}\~oes and the Department of Electrical and
Computer Engineering, Instituto Superior T\'ecnico, University of Lisbon, 1049-001, Lisboa, Portugal.
Email: Xiangrong.Zeng@lx.it.pt, mario.figueiredo@lx.it.pt.

This work was partially supported by the {\it Fundação para a Ciência e Tecnologia}, grants PEst-OE/EEI/LA0008/2013 and ...}}

\maketitle

\begin{abstract}
The ordered weighted $\ell_1$ norm (OWL) was recently proposed, with two different motivations: its good statistical properties as a sparsity promoting regularizer; the fact that it generalizes the so-called {\it octagonal shrinkage and clustering algorithm for regression} (OSCAR), which has the ability to cluster/group regression variables that are highly correlated.
This paper contains several contributions to the study and application of OWL regularization: the derivation of the atomic formulation of the OWL norm; the derivation of the dual of the OWL norm, based on its atomic formulation; a new and simpler derivation of the proximity operator of the OWL norm; an efficient scheme to compute the Euclidean projection onto an OWL ball;
the instantiation of the conditional gradient (CG, also known as Frank-Wolfe) algorithm for linear regression problems under OWL regularization; the instantiation of accelerated projected gradient algorithms for the same class of problems. Finally, a set of experiments give evidence that accelerated projected gradient algorithms are considerably faster than CG, for the class of problems considered.
\end{abstract}

\section{Introduction}

In signal processing and machine learning,  much attention has been recently devoted, not only to standard sparsity (usually enforced/encouraged by the use of an $\ell_1$ regularizer, often called LASSO \cite{tibshirani1996regression}), but also to regularizers that encourage structured/group sparsity \cite{bach2012structured}. Examples of such regularizers  include the {\it group LASSO} (gLASSO) \cite{yuan2005model}, the sparse gLASSO (sgLASSO) \cite{simon2012sparse}, the {\it fused LASSO} (fLASSO) \cite{tibshirani2004sparsity}, the {\it elastic net} (EN) \cite{zou2005regularization}, and the {\it octagonal shrinkage and clustering algorithm for regression} (OSCAR) \cite{bondell2007simultaneous} (for a more comprehensive set of references, see \cite{bach2012structured}).

The gLASSO (and its many variants and descendants \cite{bach2012structured}, \cite{Martins2011}) require the prior specification of the group structure, which is often unknown. The fLASSO, although not relying on predefined groups, depends on the order of the variables, making it unsuitable for machine learning problems, namely linear or logistic regression; in these problems, the order of the variables is usually arbitrary, thus regularizers should be invariant under permutations thereof. In contrast, both EN and the OSCAR were proposed for regression problems and are neither attached to a specific order of the variables nor to previous knowledge of the group structure.


The OSCAR regularizer (which has been shown to outperform EN in feature grouping \cite{bondell2007simultaneous,zhong2012efficient}) consists of the $\ell_1$ norm plus a sum of pairwise $\ell_\infty$ penalties, simultaneously encouraging sparsity and equality in magnitude of the estimated variables. The proximity operator of the OSCAR regularizer can be computed efficiently \cite{zeng2013solving}, \cite{zhong2012efficient}, which means that regression problems involving this regularizer can be efficiently addressed by several (accelerated) proximal gradient algorithms, such as FISTA \cite{beck2009fast}, TwIST \cite{bioucas2007new}, or SpaRSA \cite{wright2009sparse}.

A regularizer containing OSCAR, the $\ell_1$, and the $\ell_{\infty}$ norms as special cases was recently proposed \cite{bogdan2013statistical}, \cite{zeng2014decreasing}; we refer to that regularizer, which is the central object of study in this paper, as the {\it ordered weighted} $\ell_1$ (OWL) norm. Whereas in \cite{bogdan2013statistical}, the OWL norm was proposed because of its good properties in terms of controlling the {\it false discovery rate} (FDR) for variable selection with orthogonal design matrices, in \cite{zeng2014decreasing}, it was motivated as a generalization of OSCAR, for its ability to cluster/group regression variables. Very recently, the statistical performance of OWL regularization was analysed, showing its adequacy to deal with regression problems where the design matrix includes highly correlated columns \cite{FigueiredoNowak2014}.

As in the case of OSCAR, the proximity operator of the OWL norm can be computed efficiently \cite{bogdan2013statistical}, \cite{zeng2014decreasing}, with the leading cost being that of a sorting operation. This fact allows using proximal gradient algorithms ({\it e.g.}, \cite{beck2009fast}, \cite{bioucas2007new},  \cite{wright2009sparse}) to solve inverse problems or supervised learning (namely, linear regression) problems under OWL regularization in the {\it Tikhonov formulation} \cite{lorenz2013necessary}: {\it i.e.}, unconstrained optimization problems, where the objective function is the weighted sum of a loss function with a regularizer.
In this paper, we focus on an alternative approach, sometimes known as the {\it Ivanov formulation} \cite{lorenz2013necessary}, which takes the form of a constrained optimization problem, where a loss function is minimized under an upper bound constraint on the regularizer. In particular, we consider tackling this problem using either the {\it conditional gradient} (CG)  (also known as Frank-Wolfe \cite{jaggi2013revisiting}) algorithm or fast projected gradient algorithms.
The CG algorithm takes advantage of the atomic formulation of the OWL norm, which is one the contributions of this paper. The projected gradient algorithms are supported on an efficient method to compute the Euclidean projection onto a ball of the OWL norm, which is another contribution of this paper.

The main contributions of this paper are the following.
\begin{itemize}
\item Derivation of the {\it atomic norm} \cite{chandrasekaran2012convex} formulation of the OWL regularizer.
\item Derivation of the dual of the OWL norm, taking advantage of its atomic formulation.
\item Instantiation of the CG algorithm to handle the Ivanov formulation of  OWL regularization; more specifically:
\begin{itemize}
\item we show how the atomic formulation of the OWL norm allows solving efficiently the linear programming problem in each iteration of the CG algorithm;
\item based on results from \cite{jaggi2013revisiting}, we show convergence of the resulting algorithm and provide explicit values for the constants.
\end{itemize}
\item A new derivation of the proximity operator of the OWL norm, arguably simpler than those in  \cite{bogdan2013statistical} and \cite{zeng2014decreasing}, highlighting its connection to {\it isotonic regression}  and the {\it pool adjacent violators} (PAV) algorithm \cite{barlow72}, \cite{BestChakravarti}.
\item An efficient method to project onto an OWL norm ball, based on a root-finding scheme.
\item Tackling the Ivanov formulation under OWL regularization using projected gradient algorithms, based on the proposed OWL projection.
\end{itemize}

The  paper is organized as follows. Section~\ref{sec:OWL_et_al}, after reviewing the OWL norm and some of its basic properties, presents the atomic formulation, and derives its dual norm. The two key computational tools for using OWL regularization,
the proximity operator and the projection on a ball, are addressed in Section~\ref{sec:tools_owl}.
Section~\ref{sec:problems} instantiates the CG algorithm and accelerated projected gradient algorithms to tackle the constrained optimization formulation of OWL regularization for linear regression.
Finally, Section~\ref{sec:experiments} reports experimental results illustrating the performance and comparison of the proposed approaches, and Section~\ref{sec:conclusions} concludes the paper.

\subsection*{Notation}
Lower-case bold letters, {\it e.g.}, $\bf x$, $\bf y$, denote (column) vectors, their transposes are ${\bf x}^T$, ${\bf y}^T$, and the $i$-th and $j$-th components are written as $x_i$ and $y_j$. Matrices are written in upper case bold, {\it e.g.}, $\bf A$, $\bf B$.  The vector with the absolute values of the components of $\bf x$ is written as $|{\bf x}|$. For a vector $\bf x$, $x_{[i]}$ is its $i$-th largest component ({\it i.e.}, for ${\bf x}\in \mathbb{R}^n$,  $x_{[1]} \geq x_{[2]} \geq \cdots \geq x_{[n]}$, with ties broken by some arbitrary rule); consequently, $|x|_{[i]}$ is the $i$-th largest component of ${\bf x}$ in magnitude. The vector obtained by sorting (in non-increasing order) the components  of ${\bf x}$ is denoted as ${\bf x}_{\downarrow}$, thus $|{\bf x}|_{\downarrow}$ denotes the vector obtained by sorting the components of ${\bf x}$ in non-increasing order of magnitude. We denote as ${\bf P}({\bf x})$  a permutation matrix (thus ${\bf P}({\bf x})^{-1} = {\bf P}({\bf x})^{T}$) that sorts the components of $\bf x$ in non-increasing order, {\it i.e.}, ${\bf x}_{\downarrow} = {\bf P}({\bf x})\, \bf x$; naturally,  $|{\bf x}|_{\downarrow} = {\bf P}(|{\bf x}|) |{\bf x}|$. Finally, $\bone$ is a vector with all entries equal to 1, whereas $\bzero = 0\bone$ is a vector with all entries equal to zero, and $\odot$ is the entry-wise (Hadamard) product.

\begin{figure}[hbt]
  \centering
  \includegraphics[width=0.99\columnwidth]{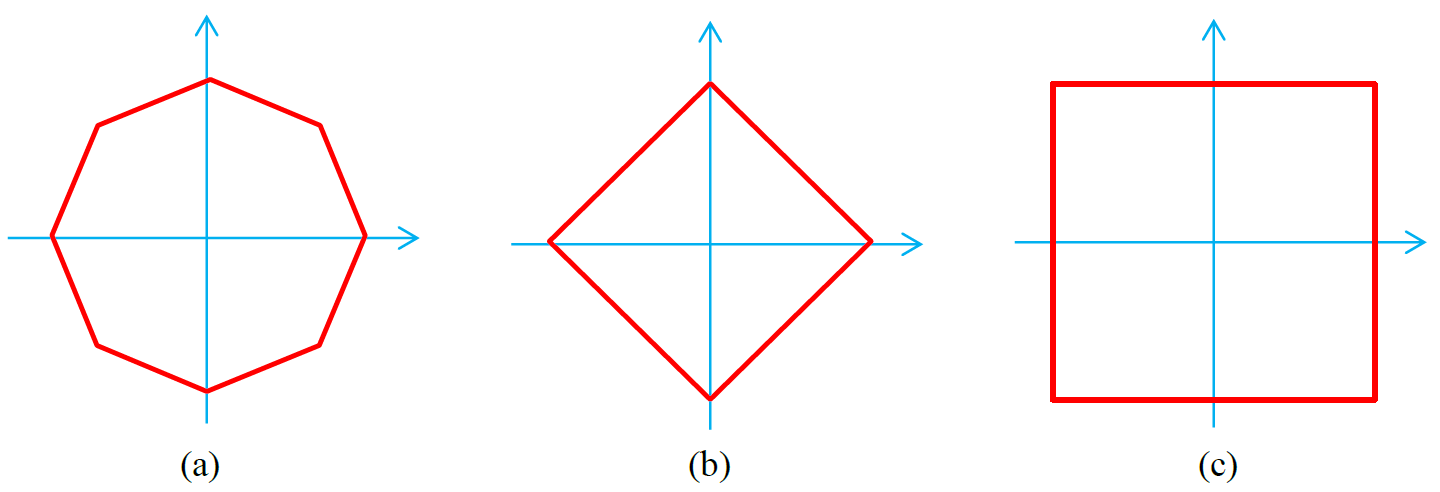}\\
  \caption{OWL balls in $\mathbb{R}^2$ with different weights: (a) $w_1 > w_2 > 0$; (b) $w_1 = w_2 > 0$; (c) $w_1 > w_2 = 0$.}\label{fig:convex_polytopes_n_2}
\end{figure}

\section{The OWL, Its Atomic Formulation, and Its Dual}\label{sec:OWL_et_al}
\subsection{The OWL Norm}\label{sec:WSL1}
\begin{figure}[htb]
  \centering
  \includegraphics[width=0.9\columnwidth]{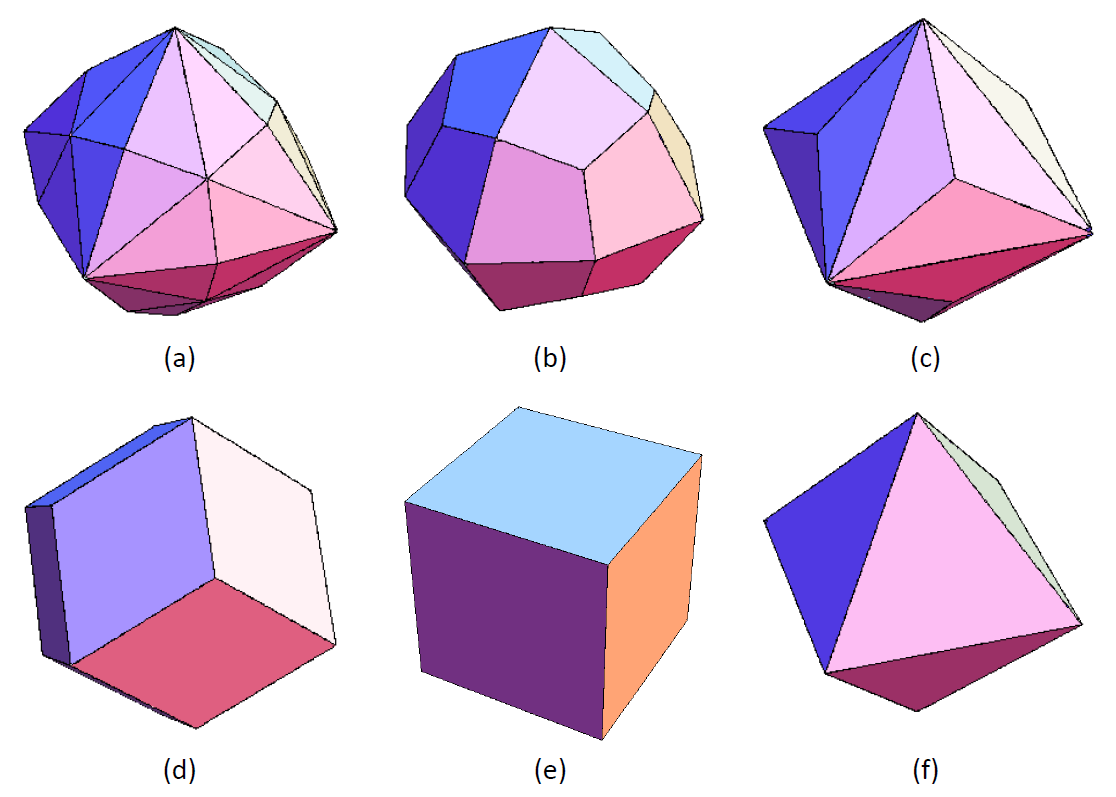}\\
  \caption{OWL balls in $\mathbb{R}^3$ with different weights: (a) $w_1 > w_2 > w_3 > 0$; (b) $w_1 > w_2 = w_3 > 0$;  (c) $w_1 = w_2 > w_3 > 0$; (d) $w_1 = w_2 > w_3 = 0$; (e) $w_1 > w_2 = w_3 = 0$; (f) $w_1 = w_2 = w_3 > 0$.}
  \label{fig:convex_polytopes_n_3}
\end{figure}

The {\it ordered weighted} $\ell_1$ (OWL) norm \cite{bogdan2013statistical}, \cite{zeng2014decreasing}, denoted as $\Omega_{\bf w}:\mathbb{R}^n \rightarrow \mathbb{R}_+$, is defined as
\begin{align}\label{eq:dwsl1}
\Omega_{{\bf w}}({\bf x}) & = \sum_{i=1}^n |x|_{[i]} \; w_i = {\bf w}^T |{\bf x}|_{\downarrow},
\end{align}
where ${\bf w}\in \mathcal{K}_{m+}$ is a vector of non-increasing weights, {\it i.e.}, belonging to the so-called {\it monotone non-negative cone} \cite{BoydVandenberghe},
\begin{equation}\label{eq:setT}
\mathcal{K}_{m+} = \{{\bf x}\in \mathbb{R}^n:\; x_1 \geq x_2 \geq \cdots x_n \geq 0\} \subset \mathbb{R}_+^n.
\end{equation}
OWL balls in $\mathbb{R}^2$ and $\mathbb{R}^3$, for different choices of the weight vector $\bf w$, are illustrated in Figs.~\ref{fig:convex_polytopes_n_2} and \ref{fig:convex_polytopes_n_3}.

The fact that, if ${\bf w}\in \mathcal{K}_{m+}\setminus \{\bzero \}$, then $\Omega_{{\bf w}}$ is indeed a norm (thus convex and homogenous of degree 1), was shown in \cite{bogdan2013statistical}, \cite{zeng2014decreasing}. It is clear that $\Omega_{{\bf w}}$ is lower bounded by the (appropriately scaled) $\ell_{\infty}$ norm:
\begin{equation}\label{eq:ineq_l_infinity}
\Omega_{{\bf w}}({\bf x}) \geq w_1 |x|_{[1]} = w_1\, \|{\bf x}\|_{\infty},
\end{equation}
with the inequality becoming an equality if $w_1 = 1$, and $w_2 = \cdots = w_n = 0$. Moreover, $\Omega_{{\bf w}}$ satisfies the following pair of inequalities, with respect to the $\ell_1$ norm:
\begin{equation}\label{eq:ineq_bar_w}
\bar{w}\, \|{\bf x}\|_1 \leq \Omega_{{\bf w}}({\bf x}) \leq w_1\, \|{\bf x}\|_1,
\end{equation}
where $\bar{w} = \|{\bf w}\|_1/n$ is the average of the weights; the first inequality is a
corollary of {\it Chebyshev's sum inequality}\footnote{According to Chebyshev's sum inequality \cite{HardyLittlewoodPolya}, if ${\bf x}, {\bf y} \in \mathcal{K}_{m+}$, then
\[
\frac{1}{n}{\bf x}^T{\bf y} \geq \left( \frac{1}{n}\sum_{i=1}^n x_i\right) \left( \frac{1}{n}\sum_{i=1}^n y_i\right) = \frac{1}{n^2} \|{\bf x}\|_1 \|{\bf y}\|_1.
\]}, and the second inequality is trivial from the definition of $\Omega_{{\bf w}}$. Of course, if ${\bf w}=w_1\bone$, both inequalities in \eqref{eq:ineq_bar_w} become equalities.

Finally, as shown in \cite{zhong2012efficient}, the following specific choice for the weights,
\begin{equation}\label{w_oscar}
{w}_i = \lambda_1 + \lambda_2 (n-i), \;\;\;\;\mbox{for $i=1,...,n$,}
\end{equation}
where $\lambda_1, \lambda_2$ are non-negative parameters,
makes the OWL norm become the OSCAR regularizer, {\it i.e.},
\begin{equation}
\Omega_{{\bf w}}({\bf x}) = \lambda_1 \|{\bf x}\|_1 + \lambda_2 \sum_{i < j} \max\{|x_i|,|x_j|\}. \label{eq:oscar}
\end{equation}
showing that the OSCAR regularizer is a particular case of the OWL norm \cite{zeng2014decreasing}.

\subsection{Atomic Norms}\label{sec:atomic_norms}
Consider a set $\mathcal{A} \subset \mathbb{R}^n$ (the collection of so-called {\it atoms}), which  is compact, centrally symmetric about the origin ({\it i.e.}, ${\bf a}\in \mathcal{A} \; \Rightarrow \; -{\bf a}\in \mathcal{A}$),
and such that $\conv(\mathcal{A})$ contains a ball of radius $\epsilon$ around the origin, for some $\epsilon > 0$ \cite{chandrasekaran2012convex}.
Then, the {\it atomic norm} of some ${\bf x}\in \mathbb{R}^n$ induced by $\mathcal{A}$ is
defined as
\begin{eqnarray}
\left\| {\bf x} \right\|_{\mathcal{A}} & = & \inf \left\{ t\geq 0:  {\bf x} \in t \: \mbox{conv}(\mathcal{A}) \right\}  \label{atomic_norm}\\
& = & \gamma({\bf x}|\mbox{conv}(\mathcal{A})),
\end{eqnarray}
where $\gamma( \cdot | C)$ is the {\it gauge} function of a convex set $C$, defined as
$\gamma({\bf x}|C) = \inf \left\{ t\geq 0:  {\bf x} \in t \: C\right\}$ \cite{Rockafellar}.

For instance, taking $\mathcal{A} = \left\{\pm {\bf e}_i\right\}$ (the set of all the vectors with one component equal to $+1$ or $-1$
and all the others equal to zero, which has cardinality $|\mathcal{A}| = 2\,n$) yields $\left\| {\bf x} \right\|_{\mathcal{A}} = \|{\bf x}\|_1$, whereas for
$\mathcal{A} = \left\{-1,\, +1\right\}^n$ (which has cardinality $|\mathcal{A}| = 2^n$), we obtain $\left\| {\bf x} \right\|_{\mathcal{A}} = \|{\bf x}\|_{\infty}$ \cite{chandrasekaran2012convex}.
The $\ell_2$ norm is recovered if $\mathcal{A}$ is the (infinite) set of all unit norm vectors.

The atomic set underlying some norm is not unique: given an atomic set $\mathcal{A}$ and another set
$\mathcal{C} \not\subset \mathcal{A}$, such that $\mathcal{C} \subset \mbox{conv}(\mathcal{A})$, then $\|\cdot \|_{\mathcal{A}} = \|\cdot\|_{\mathcal{A}\cup \mathcal{C}}$. An atomic set $\mathcal{A}$ is called {\it minimal} if there is no other set $\mathcal{A}'$  strictly contained in $\mathcal{A}$ and such that $\mbox{conv}(\mathcal{A}') = \mbox{conv}(\mathcal{A})$.

Atomic norms can be defined, not only for vectors, but also for matrices and other mathematical objects, and have recently attracted considerable interest \cite{chandrasekaran2012convex}, \cite{jaggi2013revisiting}, \cite{RaoRechtNowak2012}.

\subsection{Atomic Formulation of the OWL Norm}
Let $\mathcal{A}$ be an {\it atomic set} of the OWL norm, {\it i.e.}, a set such that $\Omega_{\bf w}({\bf x}) = \|{\bf x}\|_{\mathcal{A}}$, for any ${\bf x}\in \mathbb{R}^n$. Because $\Omega_{\bf w}({\bf x}) = \Omega_{\bf w} (|{\bf x}|_{\downarrow})$, it is enough to find the subset of $\mathcal{A}$ in the monotone non-negative cone $\mathcal{K}_{m+}$ \eqref{eq:setT}.
Let $\mathcal{B} = \mathcal{A}\cap \mathcal{K}_{m+}$ be this subset of atoms, and $\mathcal{P}_{\pm}$ denote the signed permutation  group (also called {\it hyperoctahedral group}, {\it i.e.}, the set of all $n\times n$ matrices with entries in $\{0, -1,+1\}$ and such that the sum of the absolute values in each row and column is equal to 1). The complete set $\mathcal{A}$ is obtained from $\mathcal{B}$ simply by taking all the signed permutations of all the elements thereof:
\begin{equation}\label{eq:atomic_set}
\mathcal{A} = \{ {\bf Q\, b}:\; {\bf Q}\in \mathcal{P}_{\pm} , {\bf b} \in \mathcal{B}\}.
\end{equation}
As in recent work \cite{NegrinhoMartins2014}, this set can also be written using the notion of {\it orbit} of a vector ${\bf b}$ under the action of group $\mathcal{P}_{\pm}$ ({\it i.e.}, the set $\mathcal{P}_{\pm}{\bf b} = \{ {\bf Q\, b}:\; {\bf Q}\in \mathcal{P}_{\pm} \}$):
\begin{equation}\label{eq:atomic_set2}
\mathcal{A} = \bigcup_{{\bf b}\in \mathcal{B}} \mathcal{P}_{\pm}\, {\bf b}.
\end{equation}

The next theorem (the proof of which is given in Appendix A) provides the explicit list of elements of $\mathcal{B}$, such that $\mathcal{A}$, as given by \eqref{eq:atomic_set}--\eqref{eq:atomic_set2}, is indeed an atomic set of the OWL norm.

\vspace{0.3cm}

\begin{theorem} \label{atomic_sorted}
Let $\Omega_{{\bf w}}$ be as defined in \eqref{eq:dwsl1}, where ${\bf w}\in \mathcal{K}_{m+}\setminus \{\bzero\}.$ Let $\left\| \cdot \right\|_{\mathcal{A}}$ be as defined in \eqref{atomic_norm}, where $\mathcal{A}$ is given by \eqref{eq:atomic_set} (or \eqref{eq:atomic_set2}), with
\begin{equation}\label{eq:setB}
\mathcal{B} = \left\{ {\bf b}^{(1)},...,{\bf b}^{(i)},...,{\bf b}^{(n)}\right\} \subset \mathcal{K}_{m+} ,
\end{equation}
and each ${\bf b}^{(i)}$ has the form
\begin{equation}\label{eq:bi}
{\bf b}^{(i)} = \bigl[\, \underbrace{\tau_i, ..., \tau_i}_{\mbox{$i$ entries}},0,...,0  \,\, \bigr]^T
\end{equation}
with
\begin{equation}\label{eq:tau_i}
\tau_i = \biggr( \sum_{j=1}^i {w}_j\biggr)^{-1}.
\end{equation}
Then, for any ${\bf x}\in \mathbb{R}^n$, $ \left\| {\bf x} \right\|_{\mathcal{A}} = \Omega_{{\bf w}} \left({\bf x}\right)$.
\end{theorem}
\vspace{0.3cm}

\begin{figure}[h!]
\centering
		\includegraphics[width=0.65\columnwidth]{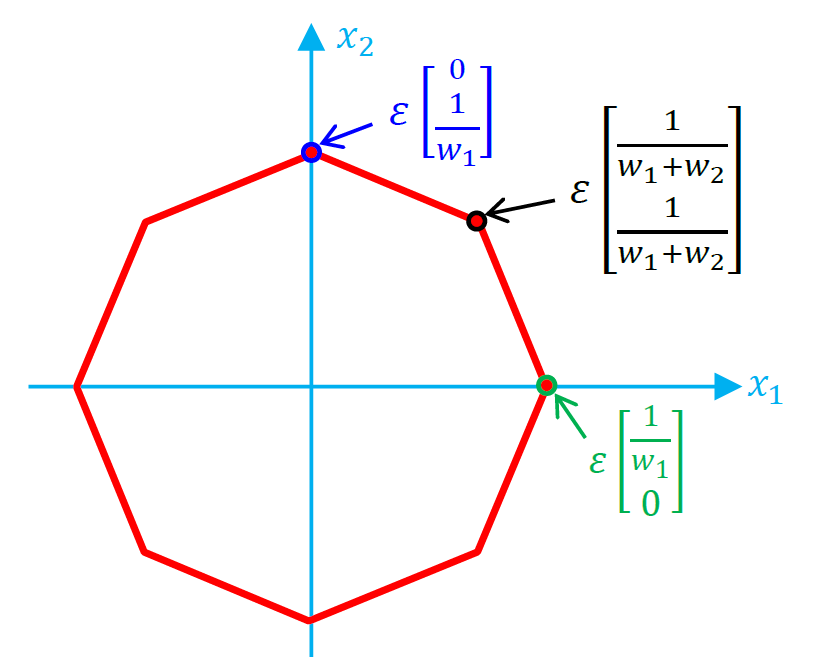}
		\caption{The atomic norm formulation of the $\varepsilon$-radius OWL ball in $\mathbb{R}^2_+$.}
		\label{fig:atomic_OWL_R2_intro}
\end{figure}

\begin{figure}[h!]
\centering
		\includegraphics[width=0.85\columnwidth]{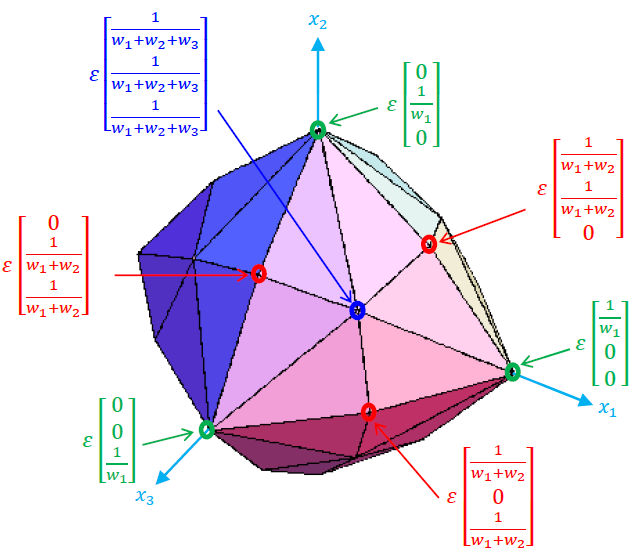}
		\caption{The atomic norm formulation of the $\varepsilon$-radius OWL ball in $\mathbb{R}^3_+$.}
		\label{fig:atomic_OWL_R3_intro}
\end{figure}

The atomic sets in $\mathbb{R}^2$ and $\mathbb{R}^3$ are represented
in Figs.~\ref{fig:atomic_OWL_R2_intro}--\ref{fig:atomic_OWL_R3_intro}.

In the general case, {\it i.e.}, if the components of ${\bf w}$ are a strictly decreasing positive sequence, the set $\mathcal{A}$ given in Theorem \ref{atomic_sorted} is a minimal atomic set (see Section \ref{sec:atomic_norms}). However, in some particular cases, $\mathcal{A}$ is not minimal. For example:
\begin{itemize}
\item if ${\bf w} = \lambda\bone$, then $\Omega_{{\bf w}} ({\bf x}) = \lambda \|{\bf x}\|_1$, for which it is known that the atomic set is simply $\mathcal{P}_{\pm} {\bf b}^{(1)}$  \cite{chandrasekaran2012convex}, \cite{NegrinhoMartins2014}. Although not minimal in this  case, the set $\mathcal{A}$ given in the theorem is still a valid atomic set, since $\mbox{conv}(\mathcal{A}) = \mbox{conv}(\mathcal{P}_{\pm} {\bf b}^{(1)})$.
\item If ${w}_1 = \lambda$ and ${w}_j = 0$, for $j=2,...,n$, then, $\Omega_{{\bf w}} ({\bf x}) = \lambda \|{\bf x}\|_{\infty}$, for which the atomic set is $\mathcal{P}_{\pm} {\bf b}^{(n)}$  \cite{chandrasekaran2012convex}, \cite{NegrinhoMartins2014}. Again, although not minimal in this case, the atomic set $\mathcal{A}$ in \eqref{eq:atomic_set}--\eqref{eq:atomic_set2} is valid, in the sense that $\mbox{conv}(\mathcal{A}) = \mbox{conv}(\mathcal{P}_{\pm} {\bf b}^{(n)})$.
\end{itemize}
The $\ell_1$ and $\ell_{\infty}$ norms are extreme examples of special cases obtained by setting a subset of the components of ${\bf w}$ to zero, or having (sub)-sequences of identical values. In those cases, although $\mathcal{A}$ may not be a minimal atomic set, it is still a valid one, and we will use the general definition in \eqref{eq:atomic_set}--\eqref{eq:atomic_set2}.

Since $\mathcal{A}$ is a finite set, $\mbox{conv}(\mathcal{A})$ is a convex polytope (see Appendix A). Moreover, it is obvious that $\mbox{conv}(\mathcal{A})$ is a full-dimensional convex polytope. In the general case (if the components of ${\bf w}$ are a strictly decreasing positive sequence), $\mathcal{A}$ is a minimal atomic set and the vertices of $\mbox{conv}(\mathcal{A})$ are the elements of $\mathcal{A}$, that is, $\mbox{vert}(\mbox{conv}(\mathcal{A})) = \mathcal{A}$. Even if $\mathcal{A}$ is not a minimal atomic set, the following inclusion is valid with generality $\mbox{vert}(\mbox{conv}(\mathcal{A})) \subseteq \mathcal{A}$ \cite{Ziegler}.

Because each ${\bf b}^{(i)}\in \mathcal{B}$ has repeated components, $|\mathcal{P}_{\pm}{\bf b}^{(i)}| < |\mathcal{P}_{\pm}| = 2^n\, n!$. For example, since any component permutation leaves ${\bf b}^{(n)}$ unchanged, $|\mathcal{P}_{\pm} {\bf b}^{(n)}| = 2^n$. In fact, it is clear that $|\mathcal{P}_{\pm}{\bf b}^{(i)} | = \binom{n}{i} 2^i$, which allows concluding that the cardinality of $\mathcal{A}$ is
\begin{equation}
|\mathcal{A}| = \sum_{i=1}^n
\binom{n}{i}\, 2^i = 3^n - 1,\label{my_A}
\end{equation}
because each orbit  $\mathcal{P}_{\pm}\, {\bf b}^{(i)}$ is disjoint from all the others. Of course, if $\mathcal{A}$ is not minimal, $|\mbox{vert}(\mbox{conv}(\mathcal{A}))| < |\mathcal{A}|$.

\subsection{Dual Norm}
Given a norm $\Omega_{{\bf w}}$, its dual norm is defined as
\begin{equation} \label{dual_norm}
\Omega^*_{{\bf w}} \left({\bf x}\right) = \|{\bf x}\|_{\mathcal{A}}^* =
\max_{\Omega_{\bf w} \left({\bf u}\right) \leq 1} {\bf u}^T {\bf x} =
\max_{\left\| {\bf u} \right\|_{\mathcal{A}} \leq 1} {\bf u}^T {\bf x}.
 \end{equation}
Using the atomic formulation, we can further write
\begin{equation}
\left\| {\bf x} \right\|_{\mathcal{A}}^* = \max_{{\bf u} \in \mbox{conv}(\mathcal{A}) }  {\bf u}^T {\bf x}\; = \; \max_{{\bf a} \in \mathcal{A} } \; {\bf a}^T {\bf x}, \label{eq:dual_last}
\end{equation}
where the second equality results from the fundamental theorem of linear programming (see, {\it e.g.}, \cite{Bertsekas2009}), according to which the maximum of a linear function over a convex polytope is attained at one (or more) of its vertices, and (as mentioned above) the vertices are contained in $\mathcal{A}$.
Using \eqref{eq:atomic_set2} and \eqref{eq:setB},
\begin{eqnarray}
\left\| {\bf x} \right\|_{\mathcal{A}}^* & = &  \max_{{\bf b}\in \mathcal{B}}  \max_{{\bf Q}\in \mathcal{P}_\pm} \;  ({\bf Q\, b})^T {\bf x} \\
& = & \max_{{\bf b}\in \mathcal{B}}  \max_{{\bf P}\in \mathcal{S}_n} \; ({\bf P\, b})^T |{\bf x}|\\
& = & \max_{{\bf b}\in \mathcal{B}}  \; {\bf b}^T |{\bf x}|_{\downarrow},\label{eq:dual_last2}
\end{eqnarray}
where $\mathcal{S}_n$ is the so-called {\it symmetric group} ({\it i.e.}, the set of all permutations of $n$ symbols), the second equality results from the fact that all the entries of ${\bf b}$ are non-negative, and the third one from the classical Hardy-Littlewood-P\'{o}lya (HLP) inequality\footnote{For any pair of vectors ${\bf x}, {\bf y} \in \mathbb{R}^n$, it holds that ${\bf x}^T {\bf y}\leq {\bf x}_{\downarrow}^T {\bf y}_{\downarrow}^{\,}$ \cite{HardyLittlewoodPolya}.} and the fact that ${\bf b}_{\downarrow} = {\bf b}$, for any ${\bf b}\in\mathcal{B}$.

Let ${\bf x}_{(i)} \in \mathbb{R}^i$ be a sub-vector of ${\bf x}\in \mathbb{R}^n$,
consisting of the $i$ largest (in magnitude) elements of ${\bf x}$; for example, $\left\| {\bf x}_{(1)}\right\|_1 =  \left\|{\bf x}\right\|_{\infty} = |x|_{[1]}$
and $\left\| {\bf x}_{(n)}\right\|_1 = \left\|{\bf x}\right\|_1$. Using this notation, and inserting in \ref{eq:dual_last} the form of the elements of $\mathcal{B}$ shown in \eqref{eq:bi}, yields the following lemma:

\vspace{0.3cm}
\begin{lemma} The dual norm of $\Omega_{\bf w}$ is given by
\begin{equation} \label{dual_norm_final}
\|{\bf x}\|_{\mathcal{A}}^* = \Omega_{\bf w}^*\left({\bf x}\right)  = \max \left\{\tau_i \bigl\|{\bf x}_{(i)}\bigr\|_1,\; i = 1, \cdots, n\right\}.
 \end{equation}
\end{lemma}


Naturally, this lemma recovers the well-known duals of the $\ell_1$ and $\ell_{\infty}$ norms. For ${\bf w} = \lambda\bone$, $\Omega_{\bf w}({\bf x}) = \lambda\|{\bf x}\|_1$,  $\tau_i = 1/(i\,\lambda)$, and the maximum in \eqref{dual_norm_final} is achieved for $i=1$, thus $\Omega_{\bf w}^*\left({\bf x}\right) =  \|{\bf x}\|_{\infty}/\lambda$. With $w_1 = \lambda$ and ${w}_j = 0$, for $j=2,...,n$, we have $\Omega_{\bf w}({\bf x}) = \lambda\|{\bf x}\|_{\infty}$ and $\tau_i = 1/\lambda$; the maximum in \eqref{dual_norm_final} is achieved for $i=n$, thus $\Omega_{\bf w}^*\left({\bf x}\right) =  \|{\bf x}\|_{1}/\lambda$.

\section{Key Computational Tools for OWL Regularization}\label{sec:tools_owl}
The key computational tools for using some norm as a regularizer are, arguably, the corresponding Moreau proximity operator \cite{Rockafellar}, \cite{bauschke2011convex}, and the Euclidean projector onto balls of that norm, which are the focus of this section.

\subsection{Proximity Operator of the OWL Norm}\label{sec:prox_OWL}
Although the proximity operator of $\Omega_{\bf w}$ (see \cite{bauschke2011convex}),
\begin{equation}\label{eq:prox_OWL}
\prox_{\Omega_{\bf w}} ({\bf v}) = \arg\min_{\bf x} \frac{1}{2}\|{\bf x - v}\|_2^2 + \Omega_{\bf w}({\bf x}),
\end{equation}
has been derived in \cite{bogdan2013statistical} and \cite{zeng2014decreasing}, we present here a new, simpler derivation, based on a recent result in \cite{Nemeth2012}. We begin with three simple lemmas (see also \cite{bogdan2013statistical}) about $\prox_{\Omega_{\bf w}}$.

\vspace{0.15cm}
\begin{lemma}\label{lem:signs}
The signs of $\prox_{\Omega_{\bf w}} ({\bf v})$ match those of ${\bf v}$,
thus $\prox_{\Omega_{\bf w}} ({\bf v}) = \sign({\bf v}) \odot  \prox_{\Omega_{\bf w}} (|{\bf v}|)$.
\end{lemma}

\vspace{0.15cm}
\begin{proof}
The lemma results from the facts that $\Omega_{\bf w}({\bf x}) = \Omega_{\bf w}(|{\bf x}|)$ and $\|{\bf v} - \sign({\bf v})\odot |{\bf x}|\|_2^2 \leq \|{\bf v - x}\|_2^2$, for any ${\bf x}$.
\end{proof}

\vspace{0.15cm}
\begin{lemma}\label{lem:sort}
The proximity operator of $\prox_{\Omega_{\bf w}} ({\bf v})$ satisfies
\begin{equation}
\prox_{\Omega_{\bf w}} ({\bf v}) = \sign({\bf v}) \odot \bigl( {\bf P}(|{\bf v}|)^T \prox_{\Omega_{\bf w}}(|{\bf v}|_{\downarrow}) \bigr).
\end{equation}
\end{lemma}

\vspace{0.1cm}
\begin{proof}
Given Lemma \ref{lem:signs}, consider, without loss of generality, that ${\bf v} \in \mathbb{R}_+^n$. The facts that $\Omega_{\bf w}({\bf x}) = \Omega_{\bf w}({\bf P\, x})$ and $\|{\bf P}({\bf v-x})\|_2^2 = \|{\bf v-x}\|_2^2$, for any permutation ${\bf P}$, immediately yield the result.
\end{proof}

\vspace{0.15cm}
\begin{lemma}\label{lem:K}
If ${\bf v}\in \mathcal{K}_{m+}$, then $\prox_{\Omega_{\bf w}} ({\bf v}) \in \mathcal{K}_{m+}$.
\end{lemma}

\vspace{0.1cm}
\begin{proof}
Write the objective function in \eqref{eq:prox_OWL} as
\begin{equation}
\frac{1}{2}\|{\bf x-v}\|_2^2 + \Omega_{\bf w}({\bf x})  =  \frac{1}{2}\|{\bf x}\|_2^2 + \frac{1}{2}\|{\bf v}\|_2^2 - {\bf v}^T{\bf x} + \Omega_{\bf w}({\bf x}), \label{eq:fromKtoK}
\end{equation}
and notice that all the terms except ${\bf v}^T{\bf x}$ are invariant under signed permutations of its arguments. The HLP inequality\footnotemark[2] guarantees that, since ${\bf v}\in \mathcal{K}_{m+}$, then ${\bf v}^T |{\bf x}|_{\downarrow} \geq {\bf v}^T {\bf x}$, for any ${\bf x}$, showing that the minimizer of \eqref{eq:fromKtoK} has to be in $\mathcal{K}_{m+}$.
\end{proof}

\vspace{0.15cm}
Lemmas \ref{lem:signs}, \ref{lem:sort}, and \ref{lem:K} show that we only need to compute
$\prox_{\Omega_{\bf w}}$ for arguments in $\mathcal{K}_{m+}$, and that the result is in $\mathcal{K}_{m+}$. Since ${\bf w},{\bf x}\in \mathcal{K}_{m+}$, thus $\Omega_{\bf w}({\bf x}) = {\bf w}^T{\bf x},$ we are  left with the problem of computing
\[
\arg\!\! \min_{{\bf x}\in \mathcal{K}_{m+}} \frac{1}{2}\|{\bf x - v}\|_2^2 + {\bf w}^T{\bf x} = \arg\!\! \min_{{\bf x}\in \mathcal{K}_{m+}} \frac{1}{2}\|{\bf x - (v-w)}\|_2^2,
\]
which is the Euclidean projection of  ${\bf (v-w)}$ onto $\mathcal{K}_{m+}$. As recently shown \cite{Nemeth2012}, the projection onto $\mathcal{K}_{m+}$ can be obtained by first computing the projection onto the monotone cone\footnote{The monotone cone is defined as
$\mathcal{K}_{m} = \{{\bf x}\in\mathbb{R}^n: \, x_1 \geq x_2 \geq \cdots \geq x_n\}$;
notice that, differently from \eqref{eq:setT}, the final inequality $x_n \geq 0$ is absent, thus $\mathcal{K}_{m}$ is not contained in the first orthant.} $\mathcal{K}_{m}$, followed by a projection onto the first orthant (simply thresholding at zero). The projection onto the monotone cone can be computed efficiently (with cost $O(n)$) by the {\it pool adjacent violators} (PAV) algorithm for isotonic regression \cite{barlow72,BestChakravarti,deLeeuw}. In summary, $\prox_{\Omega_{\bf w}}$  can be computed as
\begin{equation}
\prox_{\Omega_{\bf w}} ({\bf v}) = \sign({\bf v}) \odot \Bigl( {\bf P}(|{\bf v}|)^T \proj_{\mathbb{R}_+^n}\bigl( \proj_{\mathcal{K}_m} (|{\bf v}|_{\downarrow} - {\bf w}) \Bigr),\label{eq:prox_by_PAV}
\end{equation}
where $\proj_{\mathcal{K}_m}$ is computed via the PAV algorithm, while $\proj_{\mathbb{R}_+^n}$ is a simple clipping operation.

Interestingly, this coincides (almost) exactly with the algorithms proposed in \cite{bogdan2013statistical} and \cite{zeng2014decreasing}, although those works don't mention the connection to the PAV algorithm. Finally, it is worth pointing out that the leading computational cost of this algorithm is $O(n\log n)$, corresponding to the sorting of ${\bf v}$ into $|{\bf v}|_{\downarrow}$, since all the other operations in \eqref{eq:prox_by_PAV} have $O(n)$ cost.

\subsection{Projection Onto an OWL Ball by Root Finding}
\label{sec:owl_projection}
The projection of some ${\bf v} \in \mathbb{R}^n$ onto an $\epsilon$-radius OWL ball $\mathcal{G}_{\epsilon}^{\bf w} \equiv \{{\bf x}: \, \Omega_{\bf w}({\bf x})\leq \epsilon\}$ is given by
\begin{equation} \label{proj_atomic_ball}
\mbox{proj}_{\mathcal{G}_{\varepsilon}^{\bf w}}\! \left({\bf v}\right) = \arg\min_{{\bf x} \in \mathcal{G}_{\varepsilon}^{\bf w}} \left\|{\bf v} - {\bf x}\right\|^2_2.
\end{equation}
As far as we know, it is not possible to compute $\mbox{proj}_{\mathcal{G}_{\varepsilon}^{\bf w}}$ in closed form ({\it i.e.}, with a fixed and {\it a priori} known number of operations, as is the case of $\prox_{\Omega_{\bf w}}$ in the previous subsection), thus we consider here a root-finding-based approach, as proposed in \cite{GongGaiZhang2011,Liu2009,sra2012fast}, and which is related to efficient methods for projecting onto $\ell_1$ balls \cite{Condat2014}.

Of course, if $\Omega_{\bf w}({\bf v}) \leq \epsilon$, then $\mbox{proj}_{\mathcal{G}_{\varepsilon}^{\bf w}}\! \left({\bf v}\right)={\bf v}$, thus we will focus on the non-trivial case $\Omega_{\bf w}({\bf v}) > \epsilon$, which means that $\Omega_{\bf w}\bigl( \mbox{proj}_{\mathcal{G}_{\varepsilon}^{\bf w}}\! ({\bf v})\bigr) = \epsilon$.
The Lagrangian for \eqref{proj_atomic_ball} is
\[
L({\bf x}, \theta) = \frac{1}{2} \left\|{\bf x}-{\bf v}\right\|^2_2+ \theta \bigl( \Omega_{\bf w}({\bf x} ) - \varepsilon\bigr),
\]
where $\theta \geq 0$ is the Lagrange multiplier. Clearly, for any $\theta$, the minimizer of the Lagrangian is given by
\begin{equation}
\hat{\bf x}(\theta) = \arg\min_{\bf x} L({\bf x}, \theta) = \prox_{\, \theta\, \Omega_{\bf w}} \left( {\bf v}\right),
\end{equation}
showing that the projection is obtained by computing the proximity operator for a certain value of $\theta$. Since \eqref{proj_atomic_ball} is a convex and strictly feasible problem, strong duality holds, thus the optimal primal solution is $\hat{\bf x}(\theta^*)$, where $\theta^*$ is the optimal value of the Lagrange multiplier, found by imposing primal feasibility, {\it i.e.}, $\Omega_{\bf w}( \hat{\bf x}(\theta^*) ) = \epsilon$.
Defining the function
\begin{equation}\label{eq:function_g}
g({\theta}) = \Omega_{\bf w} \bigl(\hat{\bf x}(\theta)\bigr) - \varepsilon,
\end{equation}
shows that the optimum $\theta^*$ is a root of $g$, that is $g({\theta}^*) = 0$, suggesting that $\theta^*$ can be found by some root-finding technique \cite{GongGaiZhang2011,Liu2009,sra2012fast}. This suggestion is in fact reinforced by the following lemma (rephrased from \cite{sra2012fast})

\vspace{0.15cm}
\begin{lemma}
On the interval $[0, \Omega_{\bf w}^*({\bf v})]$, the function $g$  defined in \eqref{eq:function_g} is continuous, monotonically decreasing, and satisfies: $g(0) > 0$; for $\theta \geq \Omega_{\bf w}^*({\bf v})$, $g(\theta) = -\varepsilon < 0$. Consequently, $g$ has a unique root.
\end{lemma}
\vspace{0.15cm}

In order to find the root of $g$, we adopt the Van Wijngaarden–Dekker–Brent method \cite{Brent1973}, \cite{recipes2007} (which is used, for example, in MATLAB's  {\tt fzero} function). However, a na\"{\i}ve application of this method to function $g$ requires the repeated computation of $\prox_{\Omega_{\bf w}}$, the cost of which is dominated by a sorting operation. The following lemma (the proof of which is almost identical to those of Lemmas \ref{lem:signs} and \ref{lem:sort}, thus we omit it), together with Lemma \ref{lem:K} (which guarantees that the proximity operator does not ``unsort" its argument) opens the door to a more efficient scheme, which requires only one sorting.

\vspace{0.15cm}
\begin{lemma}\label{lem:sort_proj}
The projection $\proj_{\mathcal{G}_{\varepsilon}^{\bf w}} ({\bf v})$ satisfies
\begin{equation}
\proj_{\mathcal{G}_{\varepsilon}^{\bf w}} ({\bf v}) = \sign({\bf v}) \odot \Bigl( {\bf P}(|{\bf v}|)^T \proj_{\mathcal{G}_{\varepsilon}^{\bf w}} (|{\bf v}|_{\downarrow}) \Bigr).
\end{equation}
\end{lemma}
This lemma is exploited by noticing that
\begin{eqnarray}
\prox_{\theta\Omega_{\bf w}}(|{\bf v}|_{\downarrow}) & = & \proj_{\mathcal{K}_{m+}} (|{\bf v}|_{\downarrow} - \theta {\bf w})\label{eq:proj_a1}\\
& = & \proj_{\mathbb{R}_+^n}\Bigl( \proj_{\mathcal{K}_m} (|{\bf v}|_{\downarrow} - \theta {\bf w}),\Bigr)\label{eq:proj_a2}
\end{eqnarray}
as is clear from \ref{eq:prox_by_PAV} and Lemma \eqref{lem:K}, because $|{\bf v}|_{\downarrow} \in \mathcal{K}_{m+}$.
The proposed algorithm is presented in Fig.~\ref{fig:proj_prox}.

Some comments about the algorithm: in line 8, the function {\tt findroot}$(g,\theta_{\mbox{\small min}},\theta_{\mbox{\small max}})$ finds a root of $g$ in $[\theta_{\mbox{\small min}},\, \theta_{\mbox{\small max}}]$ (using the above mentioned Van Wijngaarden–Dekker–Brent method), where $\theta_{\mbox{\small min}} = 0$ and $\theta_{\mbox{\small max}} = u_{[1]}/\bar{w}$ is an upper-bound on $\Omega_{\bf w}^*({\bf v})$, which results from the first inequality in \eqref{eq:ineq_bar_w}; the projector $\proj_{\mathcal{K}_{m+}}$ (lines 7 and 9) is implemented by the PAV algorithm followed by clipping at zero (see \eqref{eq:proj_a1}--\eqref{eq:proj_a2}); the leading cost of the algorithm is $O(n\log n)$, associated to the unique sorting operation in line 6; all the other steps have $O(n)$ cost.

\begin{figure}
\begin{center}
\colorbox{light}{\parbox{0.95\columnwidth}{
\begin{algorithm}{Projection onto OWL ball}{\label{alg:CGM}{}}
{Input:} ${\bf v}, {\bf w}, \varepsilon$\\
{Output:} ${\bf x} = \proj_{\mathcal{G}_{\varepsilon}^{\bf w}}({\bf v})$\\
\qif $\Omega_{\bf w}({\bf v}) \leq \varepsilon$\\
\qthen  ${\bf x} = {\bf v}$ \\
\qelse remove signs: ${\bf s} = |{\bf v}|$ \\
{sort:} ${\bf u} = {\bf P}({\bf s}) {\bf s}$\\
{define:} $g({\theta}) = {\bf w}^T \proj_{\mathcal{K}_{m+}}({\bf u} - \theta\, {\bf w})  - \varepsilon$ \\
{find root:} $\theta^* = \mbox{\tt findroot}(g,0,u_{[1]}/\bar{w})$ \\
{project:} ${\bf x} = \proj_{\mathcal{K}_{m+}}({\bf u} - \theta^* {\bf w})$\\
{unsort:} ${\bf x} = {\bf P}({\bf s})^T {\bf x}$\\
{restore signs}: ${\bf x} = \sign({\bf v})\odot {\bf x}$
\qfi\\
\qreturn ${\bf x}$
\end{algorithm} }}
\end{center}
\caption{Projeciotn onto OWL ball via root finding.}\label{fig:proj_prox}
\end{figure}

\section{Solving OWL-Regularized Problems}\label{sec:problems}
\subsection{Regularization Formulations}
There are three standard formulations to combine a regularizer (here, $\Omega_{\bf w}$) and a data-fidelity term (here, simply the least squares cost typically used in linear regression, with an $m\times n$ design matrix ${\bf H}$):
\begin{enumerate}
\item Tikhonov regularization (referred to as OWL-T)
\begin{equation}\label{dwsl1_t}
\min_{{\bf x}} \tfrac{1}{2} \left\| {\bf y}-{\bf H}{\bf x} \right\|_2^2 +
\tau \; \Omega_{\bf w} ({\bf x}) ,
\end{equation}
\item Morozov regularization (referred to as OWL-M)
\begin{equation}\label{dwsl1_m}
\min_{{\bf x}} \Omega_{\bf w} ({\bf x}),
\;\;\mbox{ s.t.} \;\; \left\| {\bf y}-{\bf H}{\bf x} \right\|_2 \leq \delta,
\end{equation}
\item Ivanov regularization (referred to as OWL-I)
\begin{equation}\label{dwsl1_i}
\min_{{\bf x}} \tfrac{1}{2}\left\| {\bf y}-{\bf H}{\bf x}  \right\|_2^2 ,
 \;\; \mbox{ s.t.} \;\; \Omega_{\bf w} ({\bf x}) \leq \varepsilon,
\end{equation}
\end{enumerate}
where $\tau$, $\delta$ ,and  $\varepsilon$  are regularization parameters.
Since they are convex, these three problems are equivalent (under mild conditions), in the sense that it is possible (though, in general as difficult as solving the problem itself) to adjust the regularization parameters such that the solutions are the same \cite{Rockafellar}. However, in practice, it may be more convenient to use one or another of these formulations, either because it is easier to adjust the corresponding parameter or because the optimization problem can be more efficiently dealt with.

The OWL-T formulation \eqref{dwsl1_t} can be addressed efficiently with {\it proximal gradient algorithms}, such as FISTA \cite{beck2009fast}, TwIST \cite{bioucas2007new}, or SpaRSA \cite{wright2009sparse}, since (as shown in Section~\ref{sec:prox_OWL}) it is possible to compute the proximity operator $\prox_{\Omega_{\bf w}}$ efficiently (with $O(n \log n)$ cost).

The OWL-M formulation \eqref{dwsl1_m} can also be addressed efficiently using $\prox_{\Omega_{\bf w}}$, via algorithms based on the {\it alternating direction method of multipliers} \cite{afonso2011augmented}, \cite{boyd2011distributed}. Alternatively, since $\Omega_{\bf w}$ is a \textit{gauge}, it may be possible to use the method in \cite{BergFriedlander2011}; we will explore this possibility in future work.

This paper  focuses on the OWL-I formulation \eqref{dwsl1_i}, showing how can be addressed using either the {\it conditional gradient} (CG) algorithm (also known as the {\it Frank-Wolfe}  algorithm \cite{frank1956algorithm}, \cite{jaggi2013revisiting}, briefly reviewed in Appendix B), or  {\it projected/proximal gradient algorithms} (namely, accelerated versions such as FISTA or SpaRSA). The key different between the two approaches is that projected gradient algorithms require computing a projection onto the ball $\mathcal{G}_{\varepsilon}^{\bf w} \equiv \{{\bf x}: \, \Omega_{\bf w}({\bf x})\leq \varepsilon\}$ at each iteration, while CG does not involve any projections, but a simpler linear problem at each iteration.

\subsection{Conditional Gradient Algorithm for OWL-I} \label{sec:CG}
The CG algorithm is particularly well suited to tackle problems where the feasible set is an atomic norm ball (as \eqref{dwsl1_i}) \cite{jaggi2013revisiting}; it is a projection-free algorithm, simply requiring the solution (at each iteration) of a linear problem of the form
\begin{equation}\label{eq:dual2}
\max_{{\bf x}\in {\cal G}_\varepsilon^{\bf w}} {\bf v}^T{\bf x},
\end{equation}
for a given ${\bf v}\in \mathbb{R}^n$ (see Appendix B), the value of which defines the so-called {\it support function} of set ${\cal G}_{\varepsilon}^{\bf w}$ \cite{bauschke2011convex}, \cite{Rockafellar}. Since ${\cal G}_\varepsilon^{\bf w}$ is the ball of a norm, the value of \eqref{eq:dual2} coincides (up to a factor, which is 1 if $\varepsilon = 1$) with the corresponding dual norm.

In the following paragraphs, we show how to instantiate a CG algorithm to address the OWL-I problem \eqref{dwsl1_i}, taking advantage of the atomic formulation of the OWL norm and of the quadratic nature of the objective function.

The OWL-I problem has the form \eqref{constr_prob} (Appendix B), with $f({\bf x}) = \tfrac{1}{2}\|{\bf y-Hx}\|_2^2$, thus  $- \nabla f ({\bf x}) = {\bf H}^T( {\bf y} - {\bf H}{\bf x} )$, and the CG algorithm for solving the OWL-I problem \eqref{dwsl1_i} is as shown in Figure~\ref{fig:CG_OWL}.
Concerning line 7 of the algorithm,  following a similar chain of reasoning as in \eqref{eq:dual_last}--\eqref{eq:dual_last2} yields
\begin{equation}
{\bf s}_{k} = \mbox{sign}({\bf g}_k) \odot \bigl(
{\bf P}(|{\bf g}_k|)^T \arg \max_{{\bf b}\in \mathcal{B} } {\bf b}^T  |{\bf g}_k|_{\downarrow} \bigr).\label{eq:arg_dual}
\end{equation}

\begin{figure}[t]
\begin{center}
\colorbox{light}{\parbox{0.9\columnwidth}{
\begin{algorithm}{Conditional Gradient for OWL-I}{\label{alg:CGM_OWL}{}}
{Input:} ${\bf H}$, ${\bf y}$, ${\bf w}$, $\varepsilon$ \\
{Output:} approximate solution of \eqref{dwsl1_i}\\
Initialization: ${\bf x}_0 \in \mathcal{G}_{\varepsilon}^{\bf w}$.\\
$k=0$\\
\qrepeat\\
     ${\bf g}_k = {\bf H}^T( {\bf y} - {\bf H}{\bf x}_k)$  (* negative gradient *)\\
     ${\displaystyle {\bf s}_{k}  = \arg\max_{{\bf s} \in \mathcal{G}_{\varepsilon}^{\bf w}} {\bf s}^T {\bf g}_k}$\\
     $\gamma_k = \arg\min_{\gamma \in [0,1]} f \bigl({\bf x}_k + \gamma \, ({\bf s}_{k} - {\bf x}_k)\bigr)$\\
	 ${\bf x}_{k+1}= {\bf x}_k + \gamma_k ({\bf s}_k - {\bf x}_{k})$ \\
	 $k = k + 1$
\quntil some stopping criterion is satisfied.\\
\qreturn ${\bf x}_k$
\end{algorithm} }}
\end{center}
\caption{Instance of the CG algorithm for problem OWL-I.}\label{fig:CG_OWL}
\end{figure}

In line 8, rather than a predefined step size, we take advantage of the fact that the optimal step size can be obtained in closed form, since the objective function is quadratic \cite{figueiredo2007gradient}. In fact, letting ${\bf d}_k = {\bf s}_k - {\bf x}_{k}$, it is trivial to show that
\begin{equation}
\gamma_k = \arg\min_{\gamma \in [0,1]} f ({\bf x}_k + \gamma \, {\bf d}_{k}) = \proj_{[0,1]}\biggl(
\frac{{\bf d}_k^T {\bf g}_k  }{\| {\bf H} {\bf d}_k\|_2^2}  \biggr),\label{eq:step}
\end{equation}
where $\proj_{[0,1]}(a) = \max\{\min\{1,a\},0\}$. As shown below, this choice has the additional benefit of providing, as a zero-cost byproduct, a duality gap that upper-bounds the accuracy of the current iterate and can be used in a stopping criterion.

The leading computational cost of line 8 of the algorithm, as implemented in \eqref{eq:step}, is $O(nm)$ associated to computing the matrix-vector products involving ${\bf H}\in \mathbb{R}^{m\times n}$. The computational cost of line 7 (given in \eqref{eq:arg_dual}) is dominated by the $O(n\log n)$ cost of the sorting operation. The total cost of each iteration of the algorithm is thus $O(n\, \max\{m,\log n\})$.

The next theorem (proved in Appendix C, as a corollary of Theorem 1 in \cite{jaggi2013revisiting}) guarantees primal convergence of this instance of CG algorithm, providing explicit values for the constants.

\vspace{0.15cm}
\begin{theorem}\label{theo_CGconvergence}
Consider problem \eqref{dwsl1_i} (with ${\bf x}^*$ denoting one of its solutions) and the CG algorithm in Figure~\ref{fig:CG_OWL}. Letting $f({\bf x}) = \frac{1}{2}\|{\bf H\,x - y}\|_2^2$, the iterates satisfy
\begin{equation}\label{eq:objective_decay}
f({\bf x}_k) - f({\bf x}^*) \leq \frac{8\, \varepsilon^2 \, L }{\bar{w}^2\, (k+2)},
\end{equation}
where $L = \lambda_{\mbox{\scriptsize max}}({\bf H}^T{\bf H})$ (the largest eigenvalue of ${\bf H}^T{\bf H}$) and $\bar{w}$ is as defined in \eqref{eq:ineq_bar_w}.
\end{theorem}
\vspace{0.2cm}

Theorem \ref{theo_CGconvergence} shows that the number of iterations required to obtain an $\epsilon$-optimal solution ({\it i.e.}, such that $f({\bf x}_k) - f({\bf x}^*) \leq \epsilon$) grows like $O(1/\epsilon)$. In some problems, it may not be easy to know the Lipschitz constant $L$, making \eqref{eq:objective_decay} useless as a stopping criterion; moreover, as shown below, the bound provided in theorem is very loose. However (as show in \cite{jaggi2013revisiting}), it is possible to define the following {\it surrogate duality gap},
\begin{equation}
g({\bf x}) = \max_{{\bf s}\in \mathcal{G}_{\varepsilon}^{\bf w}} \; ({\bf x} -{\bf s})^T \nabla f({\bf x});
\end{equation}
since $f$ is convex (thus lower bounded by its local linear approximation) and ${\bf x}^*\in \mathcal{G}_{\varepsilon}^{\bf w}$,
\begin{eqnarray}
f({\bf x}^*) & \geq & f({\bf x}) + ({\bf x}^* - {\bf x})^T \nabla f({\bf x}) \\
& \geq & f({\bf x}) + \min_{{\bf s}\in \mathcal{G}_{\varepsilon}^{\bf w}} \; ({\bf s} -{\bf x})^T \nabla f({\bf x}),
\end{eqnarray}
thus $g({\bf x}) \geq f({\bf x}) - f({\bf x}^*) $, providing a certificate for the current approximation
accuracy. At each step of the algorithm, this duality gap is given by
\begin{equation}
g({\bf x}_k) = ({\bf x}_k  - {\bf s}_k)^T \nabla f({\bf x}_k) = {\bf d}_k^T {\bf g}_k,
\end{equation}
which is precisely the numerator in \eqref{eq:step}, showing that it is obtained at no additional cost. A typical use of the duality gap as a stopping criterion is to run the algorithm until the condition $g({\bf x}_k)\leq \epsilon$ (for a given $\epsilon > 0$) is satisfied, at which point it is guaranteed that $f({\bf x}_k) - f({\bf x}^*) \leq \epsilon$.

\subsection{Accelerated Projected Gradient Algorithms for OWL-I}\label{sec:FISTA_SpaRSA}
With the OWL projection ($\mbox{proj}_{\mathcal{G}_{\varepsilon}^{\bf w}}$) addressed in subsection \ref{sec:owl_projection}, the OWL-I formulation \eqref{dwsl1_i} can be efficiently addressed by accelerated projected gradient algorithms.

\subsubsection{SpaRSA} the {\it sparse reconstruction by separable approximation} algorithm \cite{wright2009sparse} is an accelerated variant of the classical iterative shrinkage-thresholding (IST) algorithm \cite{daubechies2004iterative,figueiredo2003algorithm}, which obtains its speed from using the Barzilai-Borwein (BB) step-size selection criterion \cite{barzilai1988two}, \cite{figueiredo2007gradient}. Its application to solve the OWL-I problem leads to the algorithm shown in Fig.~\ref{fig:SpaRSA}. Lines 9 and 10 implement the BB spectral step-size selection with safeguards ({\it i.e.}, bounded to the interval $[\alpha_{\min},\alpha_{\max}]$). The acceptance criterion in line 15 guarantees that the objective function decreases (see \cite{wright2009sparse} for details). Notice that, unlike other projected gradient and proximal gradient algorithm, knowledge of $L$ (the largest eigenvalue of ${\bf H}^T{\bf H}$) is not required, due to the inner backtracking loop.

\begin{figure}
\begin{center}
\colorbox{light}{\parbox{0.9\columnwidth}{
\begin{algorithm}{SpaRSA for OWL-I}{
\label{alg:sparsa}}
{Input:} ${\bf H}$, ${\bf y}$, ${\bf w}$, $\varepsilon$ \\
{Output:} approximate solution of \eqref{dwsl1_i}\\
Parameters: $\eta > 1$, $0<\alpha_{\min}<\alpha_{\max}$\\
Initialization: $\alpha_0$, ${\bf x}_0\in \mathcal{G}_{\varepsilon }^{\bf w}$\\
     ${\bf v}_0 = {\bf x}_0 -{\bf H}^T \left({\bf H} {\bf x}_0 - {\bf y}\right)/\alpha_0$\\
     ${\bf x}_{1}  = \mbox{proj}_{\mathcal{G}_{\varepsilon }^{\bf w}}\!\left({\bf v}_{0}\right)$\\
     $k=1$\\
\qrepeat\\
		 $\hat{\alpha}_k = \dfrac{\|{\bf H}({\bf x}_k - {\bf x}_{k-1})\|_2^2}{\|{\bf x}_k - {\bf x}_{k-1}\|_2^2}$\\
		 $\alpha_k = \max\left\{\alpha_{\min}, \min\left\{\hat{\alpha}_k,\alpha_{\max}\right\}\right\}$ \\
     \qrepeat\\
          ${\bf v}_k = {\bf x}_k -{\bf H}^T \left({\bf H} {\bf x}_k - {\bf y}\right)/\alpha_k$\\
          ${\bf x}_{k+1}  = \mbox{proj}_{\mathcal{G}_{\varepsilon }^{\bf w}}\!\left({\bf v}_{k}\right)$\\
		      $\alpha_k \leftarrow \eta\, \alpha_k$
		 \quntil $\|{\bf H\, x}_{k+1} - {\bf y}\|_2 \leq \|{\bf H\, x}_{k} - {\bf y}\|_2$ \\
     $k \leftarrow k + 1$
\quntil some stopping criterion is satisfied\\
\qreturn ${\bf x}_{k}$
\end{algorithm} }}
\end{center}
\caption{Instance of the SpaRSA algorithm for problem OWL-I.}\label{fig:SpaRSA}
\end{figure}

\subsubsection{FISTA} the {\it fast iterative shrinkage-thresholding algorithm} \cite{beck2009fast} is another fast variant of the IST algorithm, where the acceleration is based on Nesterov's technique \cite{nesterov1983method}, \cite{nesterovintroductory}. Because SpaRSA does not require prior knowledge of $L$, we describe a version of FISTA with backtracking (also proposed in \cite{beck2009fast}), which also does not require knowing this parameter. The resulting instantiation of  FISTA to address OWL-I problem is as shown in
Fig.~\ref{fig:FISTA}.

\begin{figure}
\begin{center}
\colorbox{light}{\parbox{0.95\columnwidth}{
\begin{algorithm}{FISTA for OWL-I}{
\label{alg:fista}}
{Input:} ${\bf H}$, ${\bf y}$, ${\bf w}$, $\varepsilon$ \\
{Output:} approximate solution of \eqref{dwsl1_i}\\
Parameter: $\eta > 1$\\
Initialization: $\alpha_0$, ${\bf x}_0 \in \mathcal{G}_{\varepsilon }^{\bf w} $\\
$t_0=1$\\
${\bf u}_1 = {\bf x}_0$\\
$k=1$ \\
\qrepeat\\
     $\alpha_k = \alpha_{k-1}$\\
     ${\bf x}_{k}  = \mbox{proj}_{\mathcal{G}_{\varepsilon}^{\bf w}}\!\left( {\bf u}_k -{\bf H}^T \left({\bf H} {\bf u}_k  - {\bf y}\right)/\alpha_k \right)$\\
\qwhile $\|{\bf H\, x}_{k} - {\bf y}\|_2 > Q_{\alpha_k}({\bf x}_k,{\bf u}_k)$\\
\qdo $\alpha_k \leftarrow \eta\, \alpha_{k}$\\
${\bf x}_{k}  = \mbox{proj}_{\mathcal{G}_{\varepsilon}^{\bf w}}\!\left( {\bf u}_k -{\bf H}^T \left({\bf H} {\bf u}_k  - {\bf y}\right)/\alpha_k \right)$ \qend \\
$t_{k+1}=\bigl(1 + \sqrt{1 + 4 t_k^2}\bigr)/2$ \\
$ {\bf u}_{k+1} = {\bf x}_{k} + \frac{t_k - 1}{t_{k+1}} \left({\bf x}_{k} - {\bf x}_{k-1}\right)$ \\
     $k \leftarrow k + 1$
\quntil some stopping criterion is satisfied\\
\qreturn ${\bf x}_{k-1}$
\end{algorithm} }}
\end{center}
\caption{Instance of FISTA (with backtracking) for problem OWL-I. The function $Q_{\alpha}({\bf x},{\bf u})$ used in line 11 is defined as (see  \cite{beck2009fast} for details): $ Q_{\alpha}({\bf x},{\bf u}) =
\|{\bf H\, u - y} \|_2^2 + 2\, ({\bf x-u})^T{\bf H}^T({\bf Hu - y})
+ \frac{\alpha}{2}\|{\bf x-u}\|_2^2$.}\label{fig:FISTA}
\end{figure}

\section{Experiments}\label{sec:experiments}
This section reports experiments to compare the performance of CG, FISTA, and SpaRSA, in solving linear regression problems with the OWL-I regularization formulation. In particular, we focus on the OSCAR regularizer \cite{bondell2007simultaneous}, which is a particular instance of the OWL norm (see Section \ref{sec:WSL1}).
All algorithms are implemented in MATLAB and run on a
64-bit Windows-7 computer, with an Intel Core i7 3.07 GHz processor and 6.0 GB of RAM. As described below, we considered both synthetic datasets and a real dataset.

\subsection{Conditional Gradient}
\begin{figure}[hbt]
\centering
		\includegraphics[width=0.70\columnwidth]{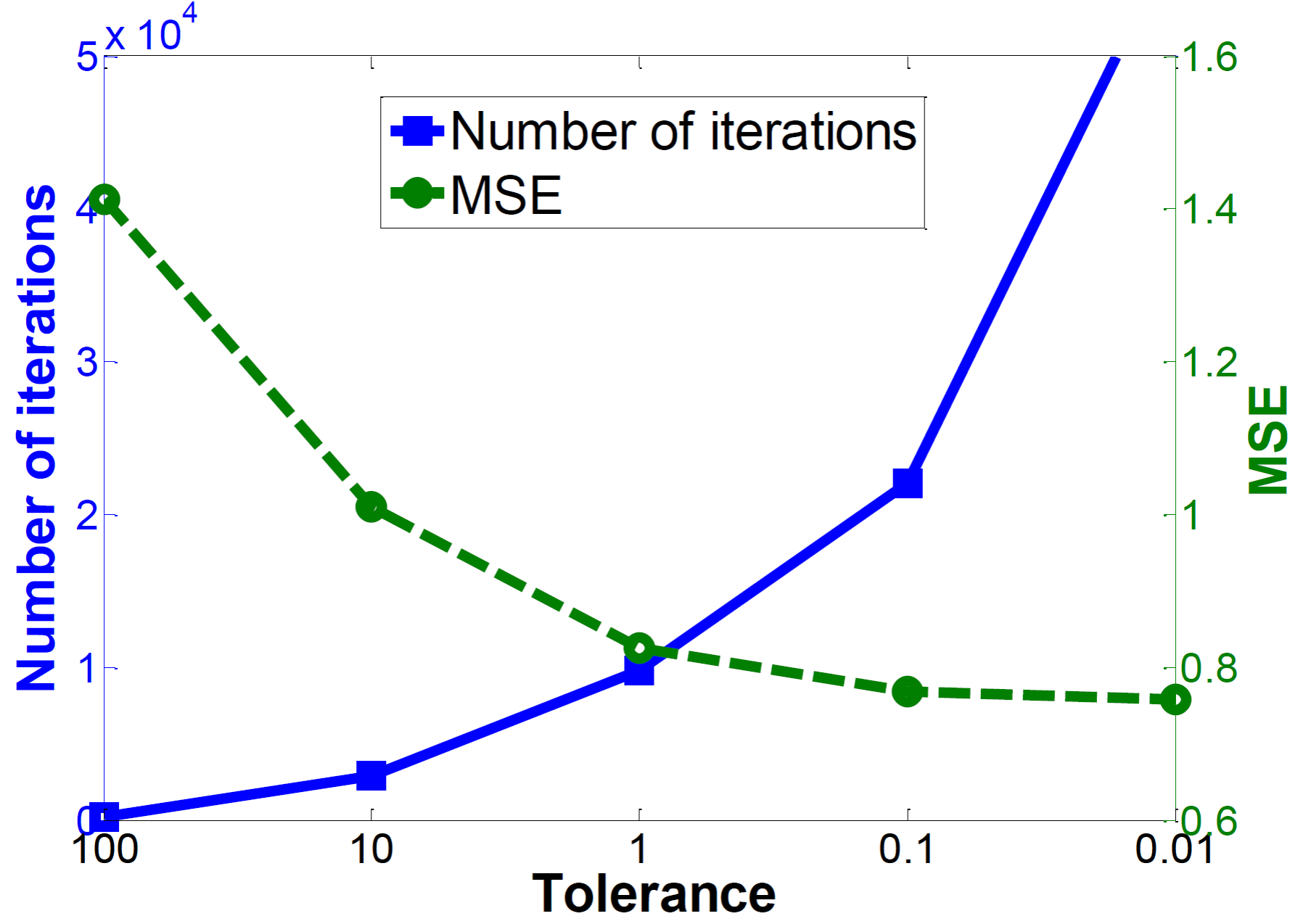}
		\caption{Evolutions of number of iterations and MSE over the tolerance $\epsilon$.
}\label{fig:exp_iter_mse}
\end{figure}

We consider a regression problem (similar to one in \cite{bondell2007simultaneous}) where the observations are generated according to ${\bf y} = {\bf H}{\bf x}_{\mbox{\small true}}  + {\bf n}$, with
${\bf H} \in \mathbb{R}^{1000 \times 1000}$ generated such that the covariance between columns $i$ and $j$ is $\mbox{cov}({\bf h}_i,{\bf h}_j) = 0.7^{\left|i-j\right|}$, then centered and standardized; the noise is Gaussian with variance $0.01$ and
\begin{multline} \label{exp_x}
{\bf x}_{\mbox{\small true}} =[\underbrace{0\cdots0}_{\tiny 150}, \underbrace{3\cdots3}_{\tiny 50}, \underbrace{0\cdots0}_{\tiny 250}, \\ \underbrace{-4\cdots-4}_{\tiny 50}, \underbrace{0\cdots0}_{\tiny 250}, \underbrace{6\cdots6}_{\tiny 50}, \underbrace{0\cdots0}_{\tiny 200}].
\end{multline}
We consider an OSCAR ($\lambda_1 = 10^{-6}$ and $\lambda_2 = 2\lambda_1$) ball of radius $\varepsilon = 1$. Following analysis in Section \ref{sec:CG}, the stopping criterion is ${\bf d}_k^T {\bf g}_k \leq \epsilon$ where ${\bf d}_k^T {\bf g}_k$ is the duality gap at $k$-th iteration, and $\epsilon$ is the tolerance.
The dependency of the number of total number of iterations and the final MSE (defined as $\left\|{\bf x}_k - {\bf x}_{\mbox{\small true}}\right\|^2_2/10^3$) with respect to $\epsilon$ is shown in
Figure~\ref{fig:exp_iter_mse}.

\begin{figure}[hbt]
\centering
		\includegraphics[width=0.70\columnwidth]{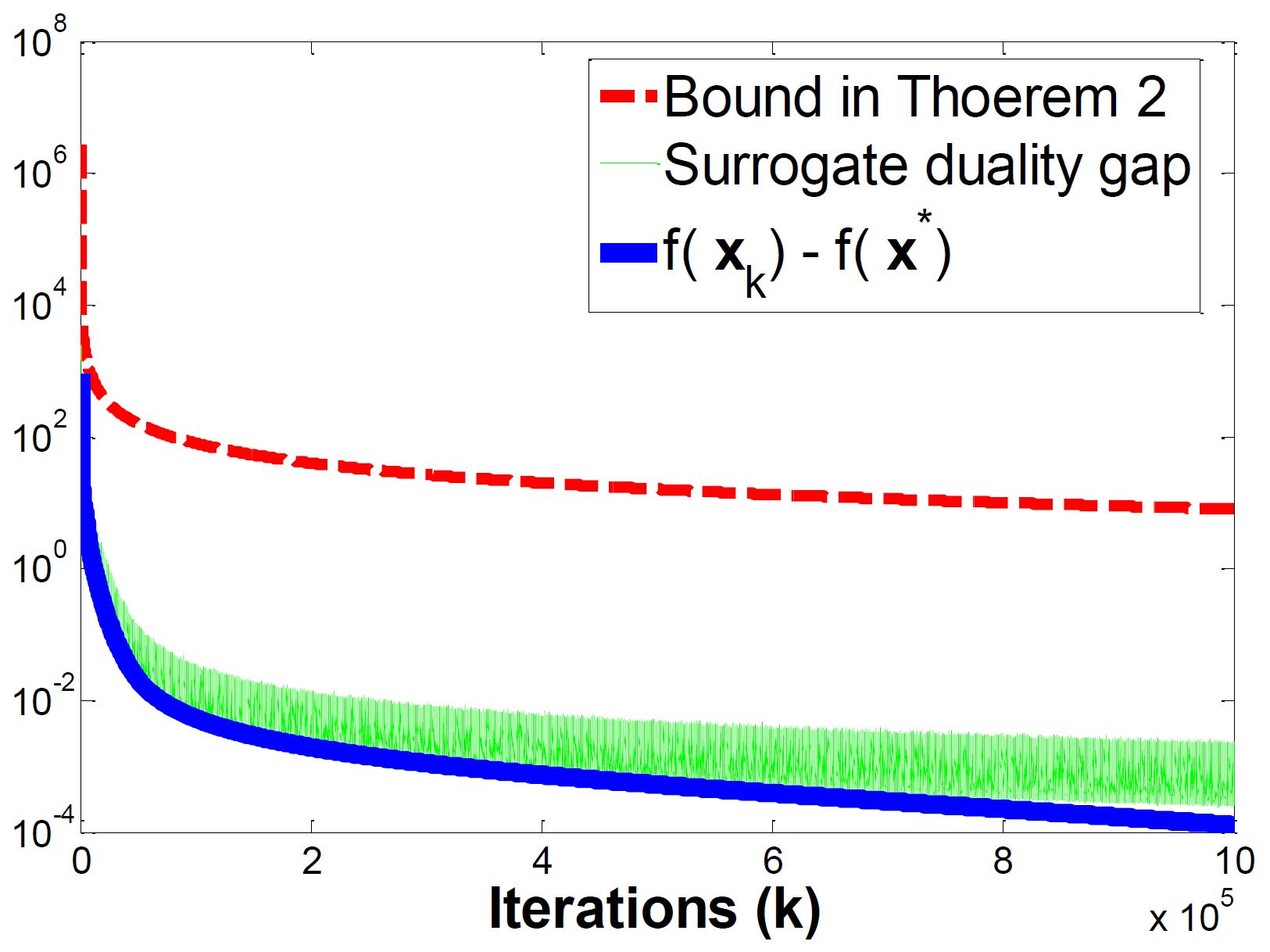}
		\caption{Evolutions of bound in Theorem \ref{theo_CGconvergence},
surrogate duality gap, and $f({\bf x}_k) - f({\bf x}^*)$, over the iterations.}
		\label{fig:bounds}
\end{figure}

The evolution of the bound in Theorem \ref{theo_CGconvergence},
the surrogate duality gap (${\bf d}_k^T {\bf g}_k$), and $f({\bf x}_k) - f({\bf x}^*)$ (where ${\bf x}^*$ is obtained when $k = 2\times 10^6$), over the iterations
are shown in Figure~\ref{fig:bounds}, from which, we can confirm that the surrogate duality gap is much tighter  than that in Theorem \ref{theo_CGconvergence}.

\subsection{CG, SpaRSA, and FISTA}
This section compares the performance of CG, FISTA (both
with and without backtracking), and SpaRSA in addressing the OWL-I formulation and the OWL-T formulation. To fairly compare these two different formulations, the experiments were conducted as follows:
\begin{itemize}
  \item Obtain an accurate estimate ${\bf x}^*$ of the OWL-T formulation, using FISTA  with a tight stopping criterion
$\left\|{\bf x}_{k+1} - {\bf x}_{k}\right\|/\left\|{\bf x}_k\right\| \leq 10^{-8}$;
\item Solve the OWL-I problem with radius $\varepsilon = \Omega_{\bf w}({\bf x}^*)$ (with this radius, the solution of the OWL-I problem will also be ${\bf x}^*$).
\end{itemize}

\begin{figure*}[hbt]
\centering
		\includegraphics[width=1.70\columnwidth]{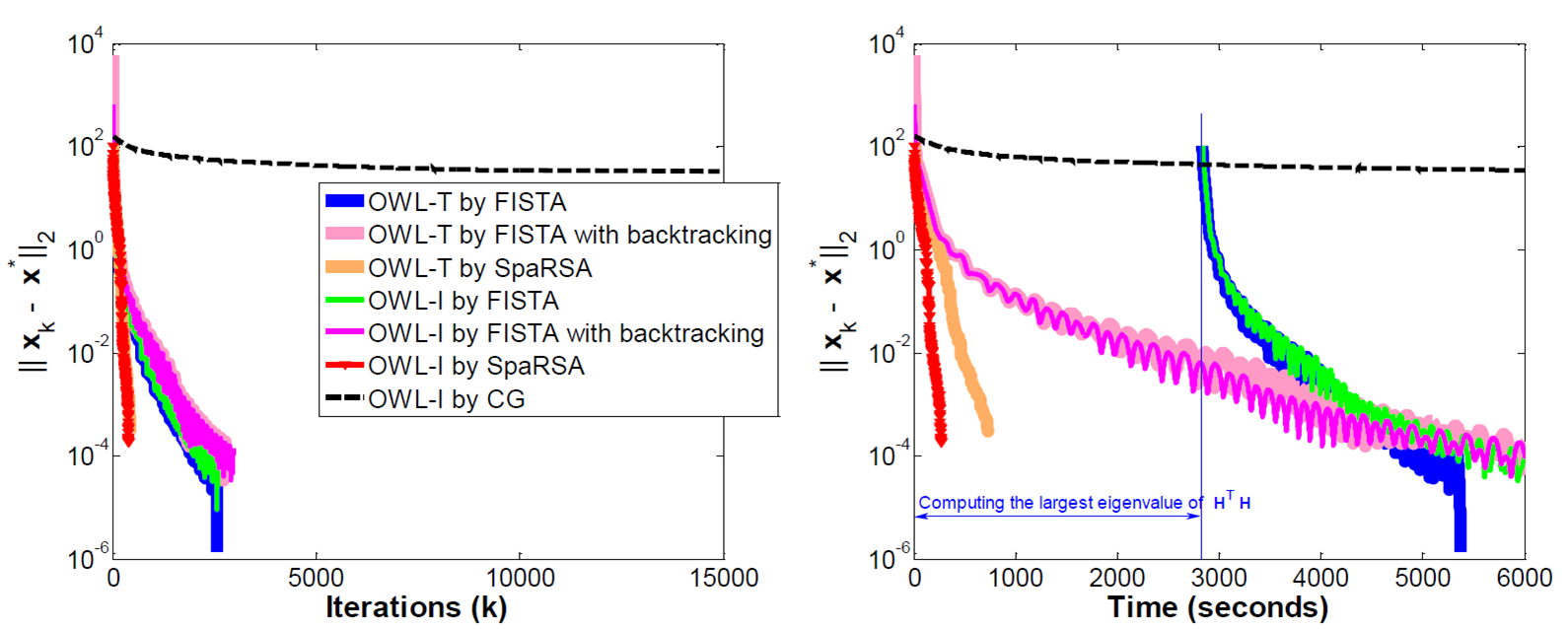}
		\caption{Evolutions of $\left\|{\bf x}_k - {\bf x}^*\right\|_2$ over iterations and time,
in case (i) with $d = 10$ (${\bf H} \in \mathbb{R}^{10000\times 10000}$).}
		\label{fig:corr_10000_10000}
\end{figure*}

\begin{figure*}[hbt]
\centering
		\includegraphics[width=1.70\columnwidth]{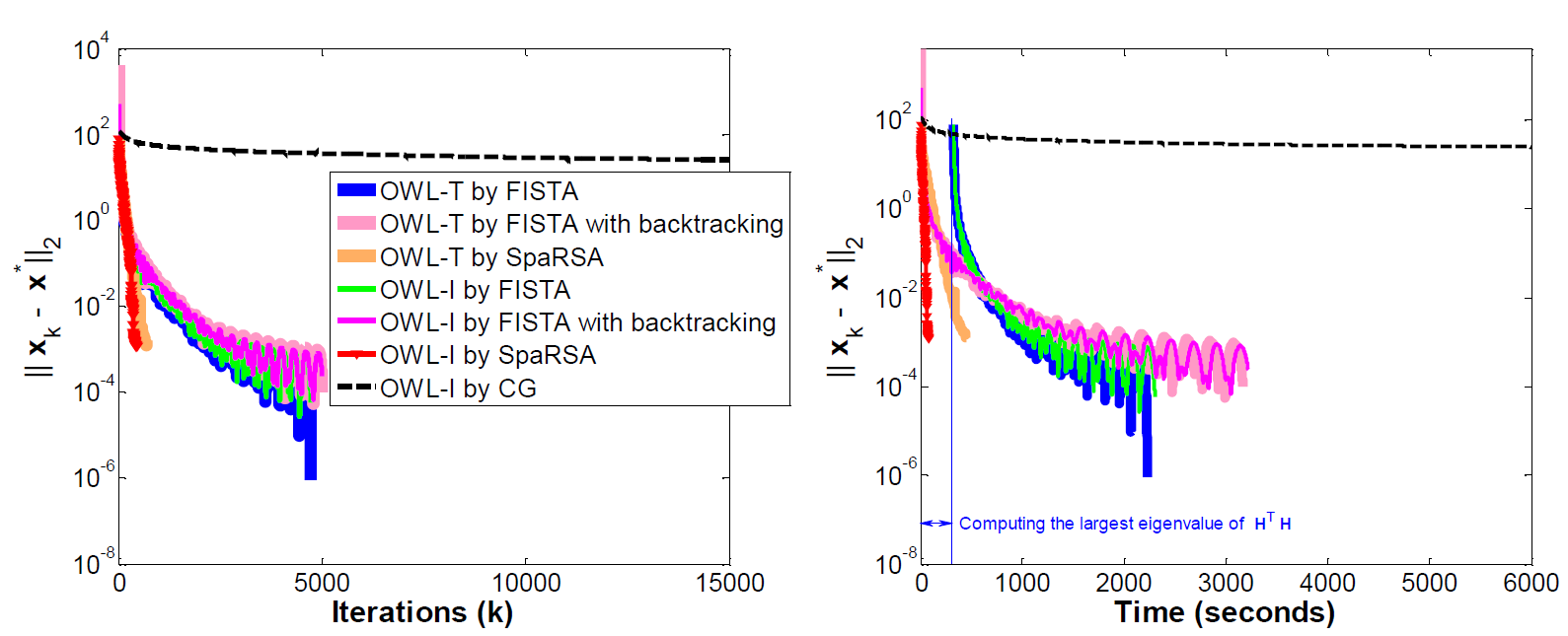}
		\caption{Evolutions of $\left\|{\bf x}_k - {\bf x}^*\right\|_2$ over iterations and time,
in case (i) with $d = 5$ (${\bf H} \in \mathbb{R}^{5000\times 5000}$).}
		\label{fig:corr_5000_5000}
\end{figure*}

\begin{figure*}[hbt]
\centering
		\includegraphics[width=1.70\columnwidth]{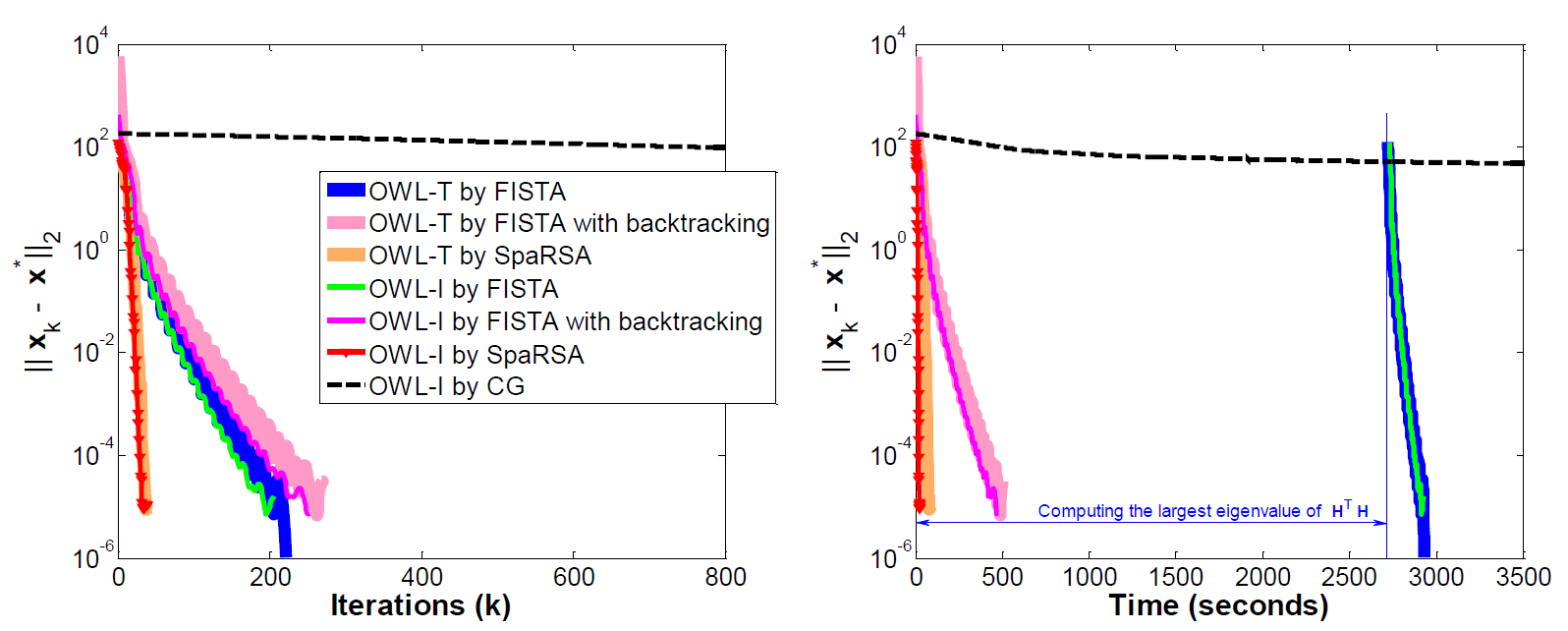}
		\caption{Evolutions of $\left\|{\bf x}_k - {\bf x}^*\right\|_2$ over iterations and time,
in case (ii) with $d = 10$ (${\bf H} \in \mathbb{R}^{10000\times 10000}$).}
		\label{fig:randn_10000_10000}
\end{figure*}

\begin{figure*}[hbt]
\centering
		\includegraphics[width=1.75\columnwidth]{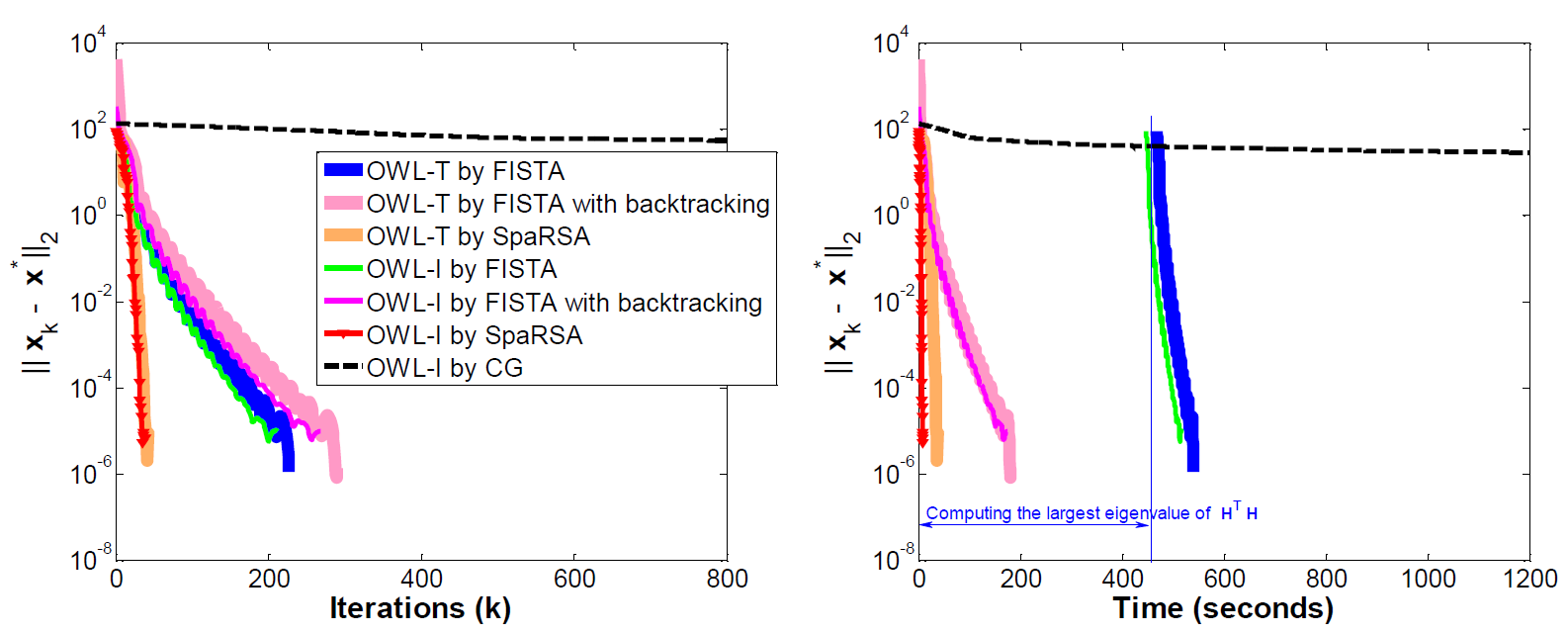}
		\caption{Evolutions of $\left\|{\bf x}_k - {\bf x}^*\right\|_2$ over iterations and time,
in case (ii) with $d = 5$ (${\bf H} \in \mathbb{R}^{5000\times 5000}$).}
		\label{fig:randn_5000_5000}
\end{figure*}

\begin{figure*}[hbt]
\centering
		\includegraphics[width=1.70\columnwidth]{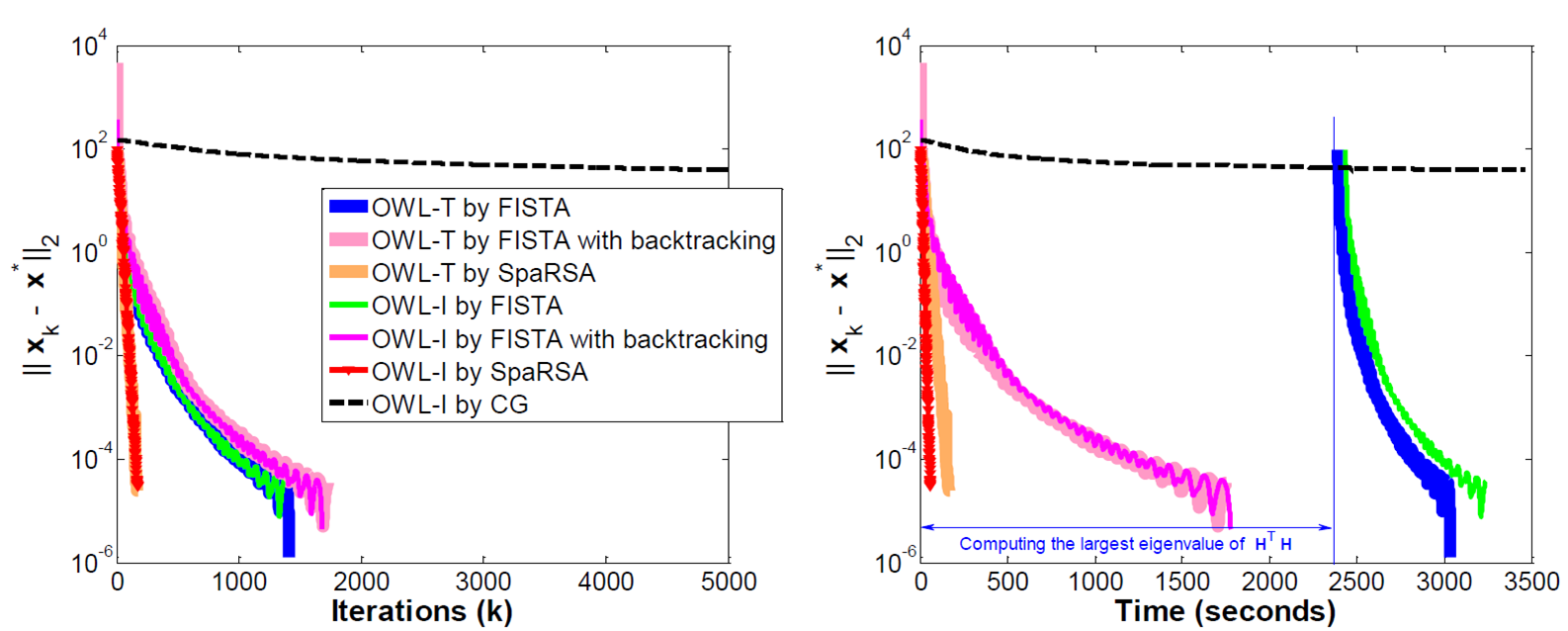}
		\caption{Evolutions of $\left\|{\bf x}_k - {\bf x}^*\right\|_2$ over iterations and time,
in case (iii) with $c=5$ and $d = 10$ (${\bf H} \in \mathbb{R}^{5000\times 10000}$).}
		\label{fig:randn_10000_5000}
\end{figure*}

\begin{figure*}[hbt]
\centering
		\includegraphics[width=1.70\columnwidth]{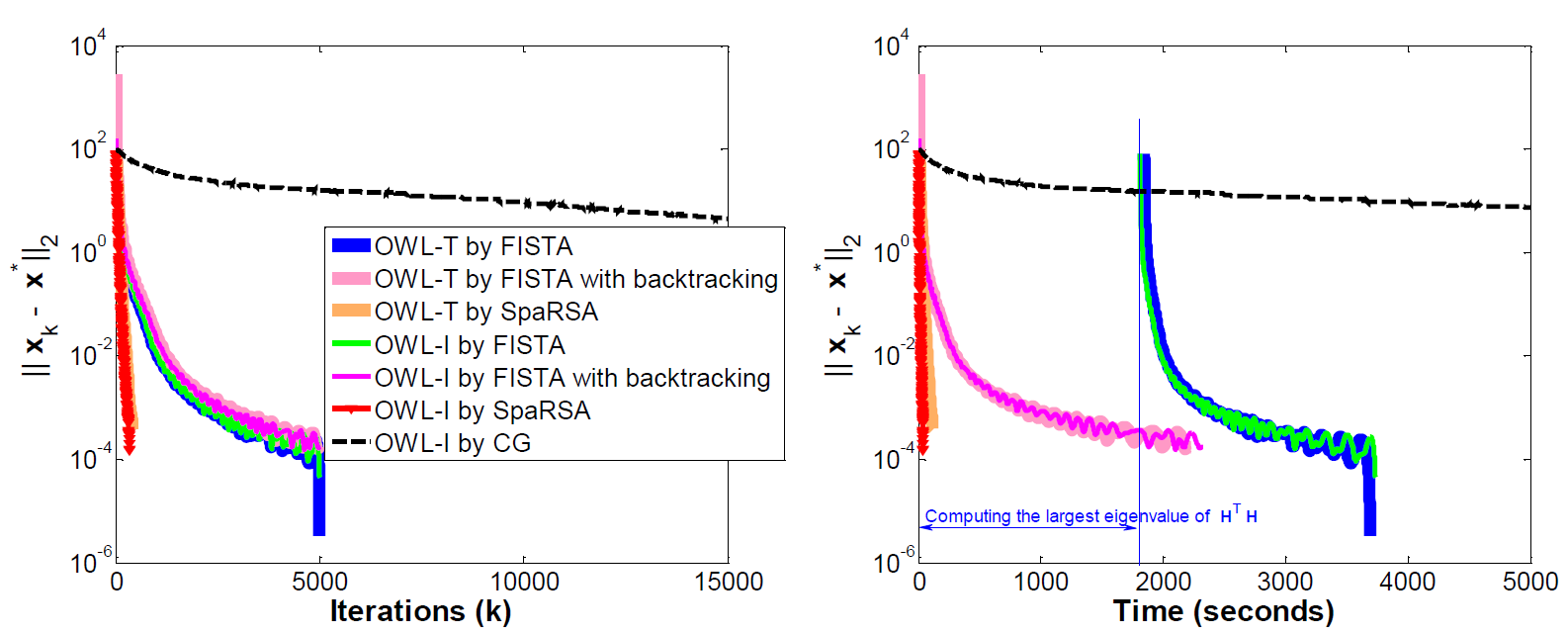}
		\caption{Evolutions of $\left\|{\bf x}_k - {\bf x}^*\right\|_2$ over iterations and time,
in case (iii) with $c=1$ and $d = 10$ (${\bf H} \in \mathbb{R}^{1000\times 10000}$).}
		\label{fig:randn_10000_1000}
\end{figure*}

We show the evolutions of $\left\|{\bf x}_k - {\bf x}^*\right\|_2$, for CG, FISTA (with and without
backtracking) and SpaRSA. CG is used to solve OWL-I,
while FISTA and SpaRSA solve both OWL-T and OWL-I. The  experimental setups are as follows. The target vector is $(1000\,d)$-dimensional ($d \in \mathbb{N}$),
\begin{multline} \label{exp_x2}
{\bf x}_{\tiny \mbox{true}} =[\underbrace{0\cdots0}_{\tiny 150d}, \underbrace{3\cdots3}_{\tiny 50d}, \underbrace{0\cdots0}_{\tiny 250d}, \\ \underbrace{-4\cdots-4}_{\tiny 50d}, \underbrace{0\cdots0}_{\tiny 250d}, \underbrace{6\cdots6}_{\tiny 50d}, \underbrace{0\cdots0}_{\tiny 200d}],
\end{multline}
and matrix ${\bf H}$ is one of the following:
\begin{description}
  \item[(i)] ${\bf H} \in \mathbb{R}^{1000d \times 1000d}$ is generated as in the previous subsection, with $d =5$ or $10$;
  \item[(ii)] ${\bf H} \in \mathbb{R}^{1000d \times 1000d}$ is sampled from a standard Gaussian,  with $d =5$ or $10$.
  \item[(iii)] ${\bf H} \in \mathbb{R}^{1000c \times 1000d}$ is sampled from a standard Gaussian,  with
 $c = 5$ or $1$, and $d =10$.
\end{description}
The noise variance is $0.01$ and we use OSCAR regularization with $\lambda_1 = 10^{-3}$ and $\lambda_2 = 10^{-5}$. The results are shown in Figures~\ref{fig:corr_10000_10000}--\ref{fig:randn_10000_1000}, from which we can observe that SpaRSA solving the OWL-I problem performs faster than the other algorithms, while FISTA performs similarly in solving the OWL-T and OWL-I problems; finally, CG is dramatically slower in these problems, although its iterations are cheaper.

Finally, we report experiments on the breast cancer dataset\footnotemark[1]\footnotetext[1]{\url{http://cbio.ensmp.fr/~ljacob/}},  which contains 8141 genes in 295 tumors, where 300 genes  are known to be most correlated with the responses. To reduce the class imbalance, we duplicate
the positive samples twice, yielding a total of 451 samples. The resulting samples are randomly split into subsets with 100, 100, and 251 samples, for  cross validation (CV) \cite{tibshirani1996regression}, training, and testing, respectively. The stopping criterion for CV and training is the same
as above with $\epsilon = 10^{-4}$, and the maximum number of iterations is set as $10^{4}$.
The total times for training and CV, as well as the test set accuracies, averaged over 50 repetitions, are shown in Table \ref{resultofbredata}, from
which, we can draw a similar conclusion as above experiments on the synthetic datasets.

\begin{table*}[t]
\centering \caption{{Results of time and test accuracy}} \label{resultofbredata}
\begin{tabular}{|l|r|r|c|}
\hline\rule[-0.1cm]{0cm}{0.4cm}
 \multirow{2}{*}{\footnotesize Algorithms} & \multicolumn{2}{|c|}{\footnotesize Time (seconds)} & {\multirow{2}{*}{\footnotesize Test accuracy }} \\
\cline{2-3}
&\footnotesize CV & \footnotesize Training & \\
\hline
\footnotesize OWL-T by FISTA                      & \footnotesize 29.1477 &  \footnotesize 0.1271   & \footnotesize 78.82 \\
\footnotesize OWL-T by FISTA with backtracking    & \footnotesize 61.3688 & \footnotesize 0.3344   & \footnotesize 78.78 \\
\footnotesize OWL-T by SpaRSA                     & \footnotesize 26.1351  & \footnotesize 0.0455    & \footnotesize 79.34   \\
\footnotesize OWL-I by FISTA                      & \footnotesize 106.0124  & \footnotesize 0.5569    & \footnotesize 79.07    \\
\footnotesize OWL-I by FISTA with backtracking    & \footnotesize 239.8232  & \footnotesize 1.4221  & \footnotesize 78.94   \\
\footnotesize OWL-I by SpaRSA                     & \footnotesize 25.6708  & \footnotesize 0.0703   & \footnotesize 79.48  \\
\footnotesize OWL-I by CG                         & \footnotesize 4237.5109  & \footnotesize 10.8428    & \footnotesize 76.76    \\
\hline
\end{tabular}
\end{table*}

\section{Conclusions}\label{sec:conclusions}
In this paper, we have made several contributions to the study of the OWL ({\it ordered weighted $\ell_1$})  norm and its use as a regularizer.
\begin{itemize}
\item We have derived the atomic formulation of the OWL norm; in addition to its potential interest for the study of this norm, the atomic formulation facilitates the use of the {\it conditional gradient} (CG) algorithm in tackling regularization problems that involve this norm.
\item Based on the atomic formulation, we have derived the dual of the OWL norm.
\item We have exploited the atomic formulation to instantiate the  CG algorithm to solve a classical constrained optimization formulation of regularized linear regression.
\item We have presented a new, arguably simpler, derivation of the proximity operator of the OWL norm, establishing its close connection to isotonic regression and the {\it pool adjacent violators} (PAV) algorithm.
\item We have shown how to efficiently compute the Euclidean projection onto a ball of the OWL norm, using a root-finding scheme.
\item We have experimentally compared CG with accelerated projected gradient algorithms, based on the proposed projection method, showing that, in the problems considered, the later are much faster than the former.
\end{itemize}

Ongoing and future work includes the application of OWL regularization to problems other than linear regression (namely, logistic regression).

\section*{Appendix A: Proof of Theorem \ref{atomic_sorted}}
Before presenting the proof, we briefly review some basic concepts of convex polytopes, which are mentioned in the paper, and a fundamental result that is used in the proof \cite{Ziegler}. A convex polytope $F \subset \mathbb{R}^n$ is the convex hull of a finite set of points $\mathcal{V} = \{{\bf v}_1,...,{\bf v}_k\} \subset \mathbb{R}^n$, that is,
\begin{equation}
F = \mbox{conv}(\mathcal{V}) = \biggl\{ \sum_{i=1}^k \lambda_i {\bf v}_i: \lambda_i \geq 0, \sum_{i=1}^k \lambda_i = 1 \biggr\}.
\end{equation}
The dimension of $F$ is that of its {\it affine hull} (the smallest affine subspace containing it), {\it i.e.}, $\dim(F) = \dim(\mbox{aff}(F))$; if $\dim(F) = n$, the polytope is called {\it full-dimensional}.
Carath\'{e}odory's Theorem states that if $F = \mbox{conv}(\mathcal{V})$ is a $p$-dimensional convex polytope, any ${\bf x}\in F$ is as a convex combination of no more than $p+1$ elements of $\mathcal{V}$.

We can now proceed to the proof of Theorem \ref{atomic_sorted}.

\proof
Since $\left\| \cdot \right\|_{\mathcal{A}} $ and $ \Omega_{{\bf w}}$ are norms, thus both homogeneous of degree 1 ({\it i.e.}, $\Omega_{\bf w}(\alpha\, {\bf x}) = |\alpha|\,\Omega_{\bf w}({\bf w})$, for any $\alpha \in \mathbb{R}$), it suffices to show that $\Omega_{\bf w}({\bf x})  = \|{\bf x}\|_{\mathcal{A}}$, for any ${\bf x}$ such that $\Omega_{\bf w}({\bf x}) = 1$. Moreover, since both are invariant under signed permutations of their arguments, {\it i.e.}, $\left\| {\bf Q\, x} \right\|_{\mathcal{A}} = \left\| {\bf x} \right\|_{\mathcal{A}}$ and $ \Omega_{{\bf w}}( {\bf Q\, x}) =  \Omega_{{\bf w}}({\bf x})$, for any ${\bf Q} \in \mathcal{P}_{\pm}$ and  ${\bf x}\in \mathbb{R}^n$, we consider, without loss of generality, that ${\bf x}\! \in\! \mathcal{K}_{m+}$ (see \eqref{eq:setT}).

We begin by showing that, if ${\bf x}\in \mathcal{K}_{m+}$ and $\Omega_{\bf w}({\bf x}) = 1$, then ${\bf x}\in \mbox{conv}(\mathcal{ B}) \subset \mbox{conv}(\mathcal{ A})$, thus $\|{\bf x}\|_{\mathcal{A}}\leq 1$. Consider the $n\times n$ matrix ${\bf B} = [{\bf b}^{(1)},\, {\bf b}^{(2)},\dots ,{\bf b}^{(n)}]$, and write ${\bf x} = {\bf B}\btheta$, where $\btheta = [\theta_1,...,\theta_n]^T$. Since ${\bf B}$ is upper-triangular with strictly positive entries, it is invertible and solving for $\btheta$ yields $\btheta = {\bf B}^{-1}{\bf x}$, where (with $x_{n+1} \equiv 0$)
\begin{equation}
\theta_i = \frac{x_i - x_{i+1}}{\tau_i}.
\end{equation}
Since ${\bf x},{\bf w}\in \mathcal{K}_{m+}\setminus \{\bzero\}$, then $x_i - x_{i+1} \geq 0$ and $\tau_i > 0$, thus $\theta_i \geq 0$. Since ${\bf x}\in \mathcal{K}_{m+}$, the condition $\Omega_{\bf w}({\bf x}) = 1$ can be written as ${\bf w}^T {\bf x} = 1$. Rearranging the corresponding sum with the {\it telescoping trick}
($\sum_{j=i}^n (x_j - x_{j+1}) = x_i$, since $x_{n+1} \equiv 0$) and noticing that $\sum_{i=1}^n  \sum_{j=i}^n  = \sum_{j=1}^n  \sum_{i=1}^j$,
\begin{eqnarray}
1\; =\; \sum_{i=1}^n w_i \, x_i & = & \sum_{i=1}^n w_i \sum_{j=i}^n (x_j - x_{j+1}) \nonumber  \\
& = & \sum_{j=1}^n  (x_j - x_{j+1}) \sum_{i=1}^j w_i \nonumber  \\
& = & \sum_{j=1}^n \frac{x_j - x_{j+1}}{\tau_j} \; = \; \sum_{j=1}^n \theta_j,
\end{eqnarray}
confirming that ${\bf x} \in \mbox{conv}(\mathcal{B})\subset \mbox{conv}(\mathcal{A})$, thus $\|{\bf x}\|_{\mathcal{A}} \leq 1$.

Having shown that, for any ${\bf x}\in \mathcal{K}_{m+}$ (thus $\Omega_{\bf w}({\bf x}) = {\bf w}^T {\bf x}$), ${\bf w}^T{\bf x} = 1$ implies that $\|{\bf x}\|_{\mathcal{A}} \leq 1$, it remains to show that it also implies that $\|{\bf x}\|_{\mathcal{A}} = 1$. Proceeding by contradiction, assume that $\|{\bf x}\|_{\mathcal{A}} < 1$, for some ${\bf x}\in \mathcal{K}_{m+}$ with ${\bf w}^T {\bf x} = 1$, which is equivalent to  ${\bf x} \in t\, \mbox{conv}(\mathcal{A})$, for some $t < 1$; from Carath\'{e}odory's Theorem, this implies that
\begin{equation}\label{eq:Carat}
{\bf x} = \sum_{i=1}^{n+1} \lambda_i {\bf Q}_i {\bf b}_i,
\end{equation}
where ${\bf Q}_i \in \mathcal{P}_{\pm}$, ${\bf b}_i \in \mathcal{B}$, $\lambda_i \geq 0$, and $\sum_{i=1}^{n+1} \lambda_i = t < 1$.

Notice now that any signed permutation matrix ${\bf Q}_i\in \mathcal{P}_{\pm}$ can be written as ${\bf Q}_i = {\bf D}_i {\bf P}_i$,  where ${\bf P}_i\in \mathcal{S}_n$ is a  permutation matrix and ${\bf D}_i = \mbox{diag}({\bf d}_i),$ with ${\bf d}_i\in \{-1,+1\}^n$, is a diagonal sign matrix. Thus, since ${\bf w},{\bf b}_i \in \mathcal{K}_{m+}$,
\begin{equation}\label{eq:ineqs}
{\bf w}^T {\bf Q}_i {\bf b}_i = {\bf w}^T {\bf D}_i {\bf P}_i {\bf b}_i \leq  {\bf w}^T {\bf P}_i {\bf b}_i \leq  {\bf w}^T {\bf b}_i = 1,
\end{equation}
where the first inequality results from both ${\bf w}$ and ${\bf P}_i {\bf b}_i$ having non-negative entries, the second one stems from the Hardy-Littlewood-P\'{o}lya inequality\footnote{For any pair of vectors ${\bf x}, {\bf y}$, it holds that ${\bf x}^T {\bf y}\leq {\bf x}_{\downarrow}^T {\bf y}_{\downarrow}^{\,}$ \cite{HardyLittlewoodPolya}.}, and ${\bf w}^T {\bf b}_i = 1$ results from the form of ${\bf b}_i \in \mathcal{B}$ (\eqref{eq:bi}--\eqref{eq:tau_i}). Combining \eqref{eq:ineqs} with \eqref{eq:Carat},
\begin{equation}
{\bf w}^T {\bf x} = \sum_{i=1}^{n+1} \lambda_i {\bf w}^T {\bf Q}_i {\bf b}_i \leq \sum_{i=1}^{n+1} \lambda_i = t < 1,
\end{equation}
contradicting that ${\bf w}^T{\bf x} =1$, thus concluding the proof.
\endproof

\section*{Appendix B: The Conditional Gradient Algorithm}
Consider a constrained convex problem of the form
\begin{equation}
 \min_{{\bf x} \in \mathbb{R}^n}  f\left({\bf x}\right)  \;  \;
 \mbox{s.t.}\;\;  {\bf x} \in \mathcal{D},\label{constr_prob}
\end{equation}
where $f$ is convex and continuously differentiable and $\mathcal{D}\neq \emptyset$
is compact and convex.

The {\it conditional gradient} (CG) is a classical method (due to Frank and Wolfe \cite{frank1956algorithm}) for problems of the form \eqref{constr_prob}, which has recently sparked a revival of interest \cite{jaggi2013revisiting}. Although there are other variants and improvements
of CG \cite{jaggi2013revisiting}, \cite{RaoShahWrightNowak2013}, we consider here the basic version presented in Figure~\ref{fig:basic_CG}. The key step of this algorithm is finding ${\bf s}_k$ (line 3), which becomes particularly convenient when ${\cal D}$ is an atomic norm ball \cite{jaggi2013revisiting}, and is in general much simpler than computing Euclidean projections onto ${\cal D}$, as required by projected gradient algorithms.

\begin{figure}
\begin{center}
\colorbox{light}{\parbox{0.8\columnwidth}{
\begin{algorithm}{Conditional Gradient}{\label{alg:CGM}{}}
Set $k=0$ and ${\bf x}_0 \in \mathcal{D}$.\\
\qrepeat\\
     ${\displaystyle {\bf s}_{k}  = \arg\min_{{\bf s} \in \mathcal{D}} {\bf s}^T \nabla f ({\bf x}_k) }$\\
     select the step-size $\gamma_k$\\
	 ${\bf x}_{k+1}= {\bf x}_k + \gamma_k ({\bf s}_{k} - {\bf x}_k)$ \\
	 $k = k + 1$
\quntil some stopping criterion is satisfied.
\end{algorithm} }}
\end{center}
\caption{The basic CG algorithm. The step-size selection procedure in line 4 may be simply a fixed expression (usually $\gamma_k = 2/(k+2)$) or some line search scheme \cite{jaggi2013revisiting}.}\label{fig:basic_CG}
\end{figure}

\section*{Appendix C: Proof of Theorem \ref{theo_CGconvergence}}
\begin{proof} Theorem \ref{theo_CGconvergence} is a corollary of the theorem in \cite{jaggi2013revisiting} that claims that the CG algorithm for a generic convex problem of the form \eqref{constr_prob} satisfies
\begin{equation}
f({\bf x}_k) - f({\bf x}^*) \leq \frac{2\, C_f }{k+2}\; (1+\delta),
\end{equation}
where $C_f$ if the so-called {\it curvature constant} of $f$ and $\delta$ is the accuracy to which the subproblems in line 3 (see Fig.~\ref{fig:basic_CG}) are solved. The algorithm in  Figure~\ref{fig:CG_OWL} uses exact solutions given by \eqref{eq:arg_dual}, thus $\delta = 0$. As also shown in \cite{jaggi2013revisiting}, if $f$ is a convex differentiable function with $L-$Lipschitz gradient with respect to some norm $\|\cdot \|$, then $C_f \leq L\, \mbox{diam}_{\|\cdot \|}\bigl( \mathcal{D}\bigr)^2$, where
\[
\mbox{diam}_{\|\cdot \|}\bigl( \mathcal{D}\bigr) = \sup_{{\bf x},{\bf z}\in \mathcal{D}} \|{\bf x}-{\bf z}\|
\]
is the diameter of set $\mathcal{D}$ w.r.t. norm $\|\cdot \|$. Function $f({\bf x}) = \frac{1}{2}\|{\bf H\,x} - {\bf y}\|_2^2$ is of course convex and differentiable with $L-$Lipschitz gradient (w.r.t. the Euclidean norm), where $L = \lambda_{\mbox{\scriptsize max}}({\bf H}^T{\bf H})$. Finally, since for any ${\bf x},{\bf z}\in \mathcal{G}_{\varepsilon}^{\bf w}$, {\it i.e.}, such that $\Omega_{\bf w}({\bf x})\leq \varepsilon$ and $\Omega_{\bf w}({\bf z})\leq \varepsilon$,
\[
 \|{\bf x}-{\bf z}\|_2 \leq \|{\bf x}-{\bf z}\|_1 \leq \frac{\Omega_{\bf w}({\bf x}-{\bf z})}{\bar{w}} \leq \frac{2\varepsilon}{\bar{w}},
\]
where the first inequality is a standard result, the second one is \eqref{eq:ineq_bar_w}, and the third is simply the triangle inequality for the norm $\Omega_{\bf w}$, we conclude that $\mbox{diam}_{\|\cdot \|}\bigl( \mathcal{G}_{\varepsilon}^{\bf w}\bigr) \leq \tfrac{2\varepsilon}{\bar{w}}$.
\end{proof}

\bibliographystyle{IEEEtranS}
\bibliography{bibfile}

\begin{thebibliography}{10}
\providecommand{\url}[1]{#1}
\csname url@samestyle\endcsname
\providecommand{\newblock}{\relax}
\providecommand{\bibinfo}[2]{#2}
\providecommand{\BIBentrySTDinterwordspacing}{\spaceskip=0pt\relax}
\providecommand{\BIBentryALTinterwordstretchfactor}{4}
\providecommand{\BIBentryALTinterwordspacing}{\spaceskip=\fontdimen2\font plus
\BIBentryALTinterwordstretchfactor\fontdimen3\font minus
  \fontdimen4\font\relax}
\providecommand{\BIBforeignlanguage}[2]{{%
\expandafter\ifx\csname l@#1\endcsname\relax
\typeout{** WARNING: IEEEtranS.bst: No hyphenation pattern has been}%
\typeout{** loaded for the language `#1'. Using the pattern for}%
\typeout{** the default language instead.}%
\else
\language=\csname l@#1\endcsname
\fi
#2}}
\providecommand{\BIBdecl}{\relax}
\BIBdecl

\bibitem{afonso2011augmented}
M.~Afonso, J.~Bioucas-Dias, and M.~Figueiredo, ``An augmented lagrangian
  approach to the constrained optimization formulation of imaging inverse
  problems,'' \emph{IEEE Transactions on Image Processing}, vol.~20, pp.
  681--695, 2011.

\bibitem{bach2012structured}
F.~Bach, R.~Jenatton, J.~Mairal, and G.~Obozinski, ``Structured sparsity
  through convex optimization,'' \emph{Statistical Science}, vol.~27, no.~4,
  pp. 450--468, 2012.

\bibitem{barlow72}
R.~Barlow, D.~Bartholomew, J.~Bremand, and H.~Brunk, \emph{Statistical
  inference under order restrictions; the theory and application of isotonic
  regression}.\hskip 1em plus 0.5em minus 0.4em\relax New York: Wiley, 1972.

\bibitem{barzilai1988two}
J.~Barzilai and J.~Borwein, ``Two-point step size gradient methods,'' \emph{IMA
  Journal of Numerical Analysis}, vol.~8, pp. 141--148, 1988.

\bibitem{bauschke2011convex}
H.~Bauschke and P.~Combettes, \emph{Convex analysis and monotone operator
  theory in Hilbert spaces}, 2011.

\bibitem{beck2009fast}
A.~Beck and M.~Teboulle, ``A fast iterative shrinkage-thresholding algorithm
  for linear inverse problems,'' \emph{SIAM Journal on Imaging Sciences},
  vol.~2, pp. 183--202, 2009.

\bibitem{Bertsekas2009}
D.~Bertsekas, \emph{Convex Optimization Theory}.\hskip 1em plus 0.5em minus
  0.4em\relax Athena Scientific, 2009.

\bibitem{BestChakravarti}
M.~Best and N.~Chakravarti, ``Active set algorithms for isotonic regression: A
  unifying framework,'' \emph{Mathematical Programming}, vol.~47, pp. 425--439,
  1990.

\bibitem{bioucas2007new}
J.~Bioucas-Dias and M.~Figueiredo, ``A new {TwIST}: two-step iterative
  shrinkage/thresholding algorithms for image restoration,'' \emph{IEEE
  Transactions on Image Processing}, vol.~16, pp. 2992--3004, 2007.

\bibitem{bogdan2013statistical}
J.~Bogdan, E.~Berg, W.~Su, and E.~Candes, ``Statistical estimation and testing
  via the ordered {$\ell_1$} norm,'' \emph{arXiv preprint
  http://arxiv.org/pdf/1310.1969v1.pdf}, 2013.

\bibitem{bondell2007simultaneous}
H.~Bondell and B.~Reich, ``Simultaneous regression shrinkage, variable
  selection, and supervised clustering of predictors with {OSCAR},''
  \emph{Biometrics}, vol.~64, pp. 115--123, 2007.

\bibitem{boyd2011distributed}
S.~Boyd, N.~Parikh, E.~Chu, B.~Peleato, and J.~Eckstein, ``Distributed
  optimization and statistical learning via the alternating direction method of
  multipliers,'' \emph{Foundations and Trends{\textregistered} in Machine
  Learning}, vol.~3, pp. 1--122, 2011.

\bibitem{BoydVandenberghe}
S.~Boyd and L.~Vandenberghe, \emph{Convex Optimization}.\hskip 1em plus 0.5em
  minus 0.4em\relax Cambridge University Press, 2004.

\bibitem{Brent1973}
R.~Brent, \emph{Algorithms for Minimization without Derivatives}.\hskip 1em
  plus 0.5em minus 0.4em\relax Prentice-Hall, 1973.

\bibitem{chandrasekaran2012convex}
V.~Chandrasekaran, B.~Recht, P.~A. Parrilo, and A.~S. Willsky, ``The convex
  geometry of linear inverse problems,'' \emph{Foundations of Computational
  Mathematics}, vol.~12, no.~6, pp. 805--849, 2012.

\bibitem{Condat2014}
\BIBentryALTinterwordspacing
L.~Condat, ``Fast projection onto the simplex and the $\ell_1$ ball,'' HAL,
  Tech. Rep. hal-01056171, 2014. [Online]. Available:
  \url{https://hal.archives-ouvertes.fr/hal-01056171}
\BIBentrySTDinterwordspacing

\bibitem{daubechies2004iterative}
I.~Daubechies, M.~Defrise, and C.~De~Mol, ``An iterative thresholding algorithm
  for linear inverse problems with a sparsity constraint,''
  \emph{Communications on pure and applied mathematics}, vol.~57, pp.
  1413--1457, 2004.

\bibitem{deLeeuw}
J.~de~Leeuw, K.~Hornik, and and, ``Isotone optimization in {R}:
  Pool-adjacent-violators algorithm (pava) and active set methods,''
  \emph{Journal of Statistical Software}, vol.~32, pp. 1--24, 2009.

\bibitem{FigueiredoNowak2014}
M.~Figueiredo and R.~Nowak, ``Sparse estimation with strongly correlated
  variables using ordered weighted $\ell_1$ regularization,'' \emph{arXiv
  preprint arXiv:}, 2014.

\bibitem{figueiredo2007gradient}
M.~Figueiredo, R.~Nowak, and S.~Wright, ``Gradient projection for sparse
  reconstruction: Application to compressed sensing and other inverse
  problems,'' \emph{IEEE Journal of Selected Topics in Signal Processing},
  vol.~1, pp. 586--597, 2007.

\bibitem{figueiredo2003algorithm}
M.~Figueiredo and R.~Nowak, ``An {EM} algorithm for wavelet-based image
  restoration,'' \emph{IEEE Transactions on Image Processing}, vol.~12, pp.
  906--916, 2003.

\bibitem{frank1956algorithm}
M.~Frank and P.~Wolfe, ``An algorithm for quadratic programming,'' \emph{Naval
  research logistics quarterly}, vol.~3, no. 1-2, pp. 95--110, 1956.

\bibitem{GongGaiZhang2011}
P.~Gong, K.~Gai, and C.~Zhang, ``Efficient {Euclidean} projections via
  piecewise root finding and its application in gradient projection,''
  \emph{Neurocomputing}, vol.~74, pp. 2754--?766, 2011.

\bibitem{HardyLittlewoodPolya}
G.~Hardy, J.~Littlewood, and G.~P\'{o}lya, \emph{Inequalities}.\hskip 1em plus
  0.5em minus 0.4em\relax Cambridge University Press, 1934.

\bibitem{jaggi2013revisiting}
M.~Jaggi, ``Revisiting {Frank-Wolfe}: Projection-free sparse convex
  optimization,'' in \emph{Proceedings of the 30th International Conference on
  Machine Learning (ICML-13)}, 2013, pp. 427--435.

\bibitem{Liu2009}
J.~Liu and J.~Ye, ``Efficient {Euclidean} projections in linear time,'' in
  \emph{Proceedings of the 26th International Conference on Machine Learning},
  2009, pp. 657--–664.

\bibitem{lorenz2013necessary}
D.~Lorenz and N.~Worliczek, ``Necessary conditions for variational
  regularization schemes,'' \emph{Inverse Problems}, vol.~29, 2013.

\bibitem{Martins2011}
A.~Martins, N.~Smith, M.~Figueiredo, and P.~Aguiar, ``Structured sparsity in
  structured prediction,'' in \emph{Conference on Empirical Methods in Natural
  Language Processing (EMNLP)}, Edinburgh, Scotland, UK, 2011.

\bibitem{NegrinhoMartins2014}
R.~Negrinho and A.~Martins, ``Orbit regularization,'' in \emph{Neural
  Information Processing Systems (NIPS) 27}, 2014.

\bibitem{nesterovintroductory}
Y.~Nesterov, ``Introductory lectures on convex optimization, 2004.''

\bibitem{nesterov1983method}
------, ``A method of solving a convex programming problem with convergence
  rate {$\mathcal{O}(1/k^2)$},'' in \emph{Soviet Mathematics Doklady}, vol.~27,
  1983, pp. 372--376.

\bibitem{Nemeth2012}
A.~Németh and S.~Németh, ``How to project on the monotone nonnegative cone
  using the pool adjacent violators type algorithms,'' available at
  \url{http://arxiv.org/pdf/1201.2343v2.pdf}, Tech. Rep., 2012.

\bibitem{recipes2007}
W.~Press, S.~Teukolsky, W.~Vetterling, and B.~Flannery, \emph{Numerical
  Recipes: The Art of Scientific Computing (3rd Edition)}.\hskip 1em plus 0.5em
  minus 0.4em\relax Cambridge University Press, 2007.

\bibitem{RaoRechtNowak2012}
N.~Rao, B.~Recht, and R.~Nowak, ``Universal measurement bounds for structured
  sparse signal recovery,'' in \emph{Proc. Intern. Conf. Artificial
  Intelligence and Statistics (AISTATS)}, 2012, pp. 942--950.

\bibitem{RaoShahWrightNowak2013}
N.~Rao, P.~Shah, S.~Wright, and R.~Nowak, ``A greedy forward backward method
  for atomic norm constrained minimization,'' in \emph{Proc. IEEE Intern. Conf.
  Acoustics, Speech and Signal Processing (ICASSP)}, 2013.

\bibitem{Rockafellar}
R.~T. Rockafellar, \emph{Convex Analysis}.\hskip 1em plus 0.5em minus
  0.4em\relax Princeton University Press, 1970.

\bibitem{simon2012sparse}
N.~Simon, J.~Friedman, T.~Hastie, and R.~Tibshirani, ``The sparse-group
  lasso,'' \emph{Journal of Computational and Graphical Statistics}, 2012, to
  appear.

\bibitem{sra2012fast}
S.~Sra, ``Fast projections onto mixed-norm balls with applications,''
  \emph{Data Mining and Knowledge Discovery}, vol.~25, no.~2, pp. 358--377,
  2012.

\bibitem{tibshirani1996regression}
R.~Tibshirani, ``Regression shrinkage and selection via the lasso,''
  \emph{Journal of the Royal Statistical Society (B)}, pp. 267--288, 1996.

\bibitem{tibshirani2004sparsity}
R.~Tibshirani, M.~Saunders, S.~Rosset, J.~Zhu, and K.~Knight, ``Sparsity and
  smoothness via the fused lasso,'' \emph{Journal of the Royal Statistical
  Society (B)}, vol.~67, pp. 91--108, 2004.

\bibitem{BergFriedlander2011}
E.~{van den Berg} and M.~Friedlander, ``Sparse optimization with least-squares
  constraints,'' \emph{SIAM Journal on Optimization}, vol.~21, pp. 1201--1229,
  2011.

\bibitem{wright2009sparse}
S.~Wright, R.~Nowak, and M.~Figueiredo, ``Sparse reconstruction by separable
  approximation,'' \emph{IEEE Transactions on Signal Processing}, vol.~57, pp.
  2479--2493, 2009.

\bibitem{yuan2005model}
M.~Yuan and Y.~Lin, ``Model selection and estimation in regression with grouped
  variables,'' \emph{Journal of the Royal Statistical Society (B)}, vol.~68,
  pp. 49--67, 2005.

\bibitem{zeng2014decreasing}
X.~Zeng and M.~Figueiredo, ``Decreasing weighted sorted $\ell_1$
  regularization,'' \emph{IEEE Signal Processing Letters}, vol.~21, pp.
  1240--1244, 2014.

\bibitem{zeng2013solving}
------, ``Solving {OSCAR} regularization problems by fast approximate proximal
  splitting algorithms,'' \emph{Digital Signal Processing}, vol.~31, pp.
  124--135, 2014.

\bibitem{zhong2012efficient}
L.~Zhong and J.~Kwok, ``Efficient sparse modeling with automatic feature
  grouping,'' \emph{IEEE Transactions on Neural Networks and Learning Systems},
  vol.~23, pp. 1436--1447, 2012.

\bibitem{Ziegler}
G.~Ziegler, \emph{Lectures of Polytopes}.\hskip 1em plus 0.5em minus
  0.4em\relax Springer, 1995.

\bibitem{zou2005regularization}
H.~Zou and T.~Hastie, ``Regularization and variable selection via the elastic
  net,'' \emph{Journal of the Royal Statistical Society (B)}, vol.~67, pp.
  301--320, 2005.

\end{thebibliography}

\end{document}